\documentclass[11pt, a4paper]{article}

\usepackage[T1]{fontenc} 
\usepackage{mathptmx}    
\usepackage{CJKutf8}

\usepackage{geometry}
\usepackage{algorithm}
\usepackage{algpseudocode}
\usepackage{graphicx}
\usepackage{amsmath, amssymb}
\usepackage{caption}
\usepackage{hyperref}
\usepackage{booktabs}
\usepackage{natbib} 
\hypersetup{colorlinks=true, allcolors=blue}
\usepackage{multirow}
\usepackage[table]{xcolor}
\usepackage{arydshln}
\usepackage{float}
\usepackage{subcaption}
\usepackage{listings}
\usepackage{enumitem}
\usepackage{wrapfig}
\usepackage{titletoc}
\usepackage{tabularx}
\usepackage{placeins}

\usepackage{fancyhdr}
\usepackage{titlesec}
\titleformat{\section}{\large\bfseries}{\thesection}{1em}{\MakeUppercase}
\titleformat{\subsection}{\normalsize\bfseries}{\thesubsection}{1em}{}
\titleformat{\subsubsection}{\normalsize\bfseries}{\thesubsubsection}{1em}{}

\title{\LARGE \textbf{MagicSelector: Joint Optimization for Agent Tool Selection via Counterfactual Decomposition and Progressive Reranking}}

\author{
  \normalsize
  \textbf{HONOR Agentic Search Team}
}
\date{} 

\begin{document}

\maketitle
\thispagestyle{empty} 

\renewcommand{\abstractname}{\large \textbf{ABSTRACT}}
\begin{abstract}
\noindent We present MagicSelector, a joint optimization framework integrating Counterfactual task decomposition, Progressive reranking, and Dynamic Top-K, designed to address the fundamental challenges of tool retrieval in agents. MagicSelector is a specialized framework capable of translating ambiguous user instructions into executable atomic subtasks and guiding high-precision tool retrieval, effectively mitigating redundant noise and severe context distraction in out-of-domain (OOD) scenarios. We empower MagicSelector with these capabilities through three key contributions: (1) a preference-guided counterfactual task decomposition mechanism that utilizes a counterfactual reward to quantify the marginal causal gain of decomposition on retrieval ranking, effectively imposing fine-grained structural supervision on logical coherence; (2) a progressive tool reranking method driven by self-distillation hard negative mining, which optimizes both point-wise and list-wise relevance to enhance fine-grained discrimination among highly similar tools; and (3) a dual semantic boundary-aware dynamic Top-K strategy that adaptively monitors reranking score cliffs and inter-tool semantic shifts to dynamically truncate the candidate list, maximizing relevant tool recall while filtering long-tail noise. Evaluated on MTDTool, the first task decomposition benchmark we constructed tailored for mobile multi-turn interactions with process-level annotations, MagicSelector yields promising performance. Extensive experiments demonstrate that MagicSelector significantly outperforms state-of-the-art methods in terms of tool retrieval accuracy, OOD generalization capability, and overall token efficiency, thereby demonstrating the effectiveness of our proposed framework.

\end{abstract}

\vspace{1cm}

\section{Introduction}

Recent advancements in Large Language Models (LLMs)~\citep{zhao2026survey,naveed2025comprehensive} have catalyzed a fundamental shift from passive text generation to the development of autonomous agents capable of interacting with external environments through tool invocation~\citep{huang2024understanding}. A prominent and highly practical instantiation of this paradigm is the mobile agent, which operates in complex, real-world reasoning environments. In such settings, user interactions are highly diverse, encompassing multi-turn contexts, ambiguous references, and complex multi-task instructions within a single utterance.

To fulfill these complex instructions, mobile agents must interact with a rapidly growing ecosystem of external tools. However, as feeding all tool descriptions into a limited context window is computationally intractable, tool retrieval~\citep{shi2025retrieval} has emerged as the critical cognitive bottleneck. The evolution of tool retrieval has primarily bifurcated into static and dynamic injection paradigms. While dynamic injection enables agents to actively fetch tools to explore open environments~\citep{qian2023creator,cai2024large,du2024anytool}, this active acquisition process inevitably increases system latency. Furthermore, as practical mobile applications increasingly adopt Model Context Protocol (MCP) tools, there is a strict requirement for recalling highly precise tools to ensure exact schema matching and execution safety. Therefore, to balance low latency with high precision, practical tool retrieval for mobile agents must predominantly rely on static knowledge injection~\citep{schick2023toolformer,patil2024gorilla,qin2024toolllm}.

Crucially, under this static retrieval paradigm, existing efforts mostly focus on straightforward, single-intent queries. Directly matching the convoluted user instructions of mobile agents against a static library of atomic tools inevitably leads to severe semantic mismatch and retrieval failure. Therefore, fine-grained task decomposition is highly necessary to act as a deep planning mechanism---translating multi-turn and multi-task dialogues into executable atomic subtasks before retrieval. However, unlike general text retrieval, constructing an effective tool retrieval pipeline driven by task decomposition inherently involves several critical challenges.

\textbf{First}, how can we prevent the decomposition model from maximizing retrieval matching through shortcut strategies, such as repetitive decomposition, during optimization? Early prompt-based and supervised fine-tuning (SFT) methods~\citep{yao2022react,brauns2025toolreagt,jagerman2023query,huang2024planning} suffer from cumulative errors. Meanwhile, recent reinforcement learning (RL) methods~\citep{fang2026beyond,guo2025deepseek} lack causal attribution in their reward signals. By relying on static, outcome-oriented metrics, models easily fall into local optima. Specifically, they tend to exploit spurious correlations between the shallow features of decomposition results and retrieval metrics, rather than performing precise decomposition. This severely impairs generalization in out-of-domain (OOD) scenarios involving unseen tools.
\textbf{Second}, how can we accurately distinguish functionally similar tools? Given the significant semantic gap between heterogeneous user queries and massive tool documentation, traditional reranking paradigms~\citep{gangi2024first,yoon2024listt5,Zhi2026CompressthenRankFA} lack deep token-level interactions. Moreover, relying on random negative samples provides weak learning signals, limiting the model's discriminative capability. 
\textbf{Third}, how should we dynamically filter noise while preserving crucial tools to maintain context efficiency? Existing retrieval mechanisms predominantly rely on static Top-K truncation heuristics. A conservative $K$ value causes false negatives for crucial tools, whereas an aggressive $K$ value inevitably introduces redundant long-tail noise, exacerbating inference overhead and inducing context distraction phenomena like the lost-in-the-middle effect.



\begin{figure}[htbp]
  \centering
  \includegraphics[width=1\textwidth]{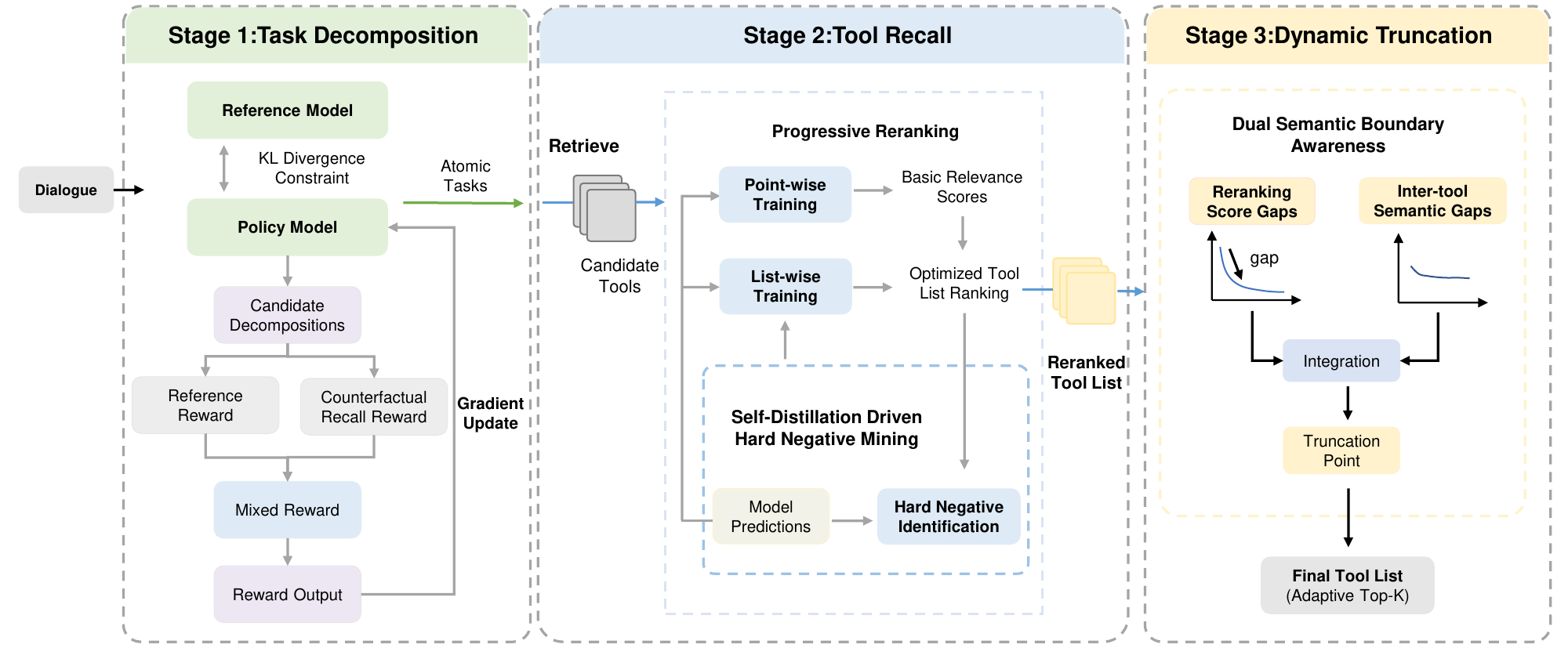}
  \caption{\textbf{The overall framework of MagicSelector.} It presents a joint optimization pipeline integrating counterfactual task decomposition, progressive reranking, and dynamic Top-K for complex tool-use scenarios. The framework consists of three key phases: \textbf{(1) Task Decomposition via Counterfactual RL}, which introduces a counterfactual reward mechanism based on relative ranking gains to provide precise causal attribution, guiding the agent to learn semantically accurate and structurally robust decomposition strategies; \textbf{(2) Progressive Tool Reranking with Hard Negative Mining}, which employs a Cross-Encoder for point-wise and list-wise optimization, coupled with self-distillation to automatically mine hard negative samples, thereby enhancing fine-grained discrimination among highly similar tools; \textbf{(3) Dual Semantic Boundary-Aware Dynamic Top-K}, which adaptively truncates the candidate list by monitoring both reranking score cliffs and inter-tool semantic shifts, effectively maximizing relevant tool recall while filtering long-tail noise to reduce token consumption and mitigate the lost-in-the-middle effect.}
  \label{fig:framework}
  \vspace{-13pt}
\end{figure}


In this work, we propose MagicSelector, a joint optimization framework designed to address these fundamental challenges by integrating counterfactual task decomposition~\citep{zhao2026pctd}, progressive reranking, and dynamic Top-K (see Figure~\ref{fig:framework}). Specifically, MagicSelector features three primary aspects corresponding to the aforementioned challenges. 
To address the first challenge, we introduce a preference-guided counterfactual task decomposition mechanism~\citep{kaddour2025causal}. By utilizing a counterfactual reward to quantify the marginal causal gain of decomposition on retrieval ranking, it shifts the optimization objective from outcome matching to process gain, effectively cutting off spurious correlations and providing an effective gradient direction for the policy model. This approach is applicable to single-turn dialogues, multi-turn dialogues, as well as single-utterance multi-task scenarios.

To tackle the second challenge, we propose a progressive tool reranking method driven by self-distillation hard negative mining. This method optimizes both point-wise and list-wise relevance, and iteratively mines high-scoring incorrect tools as hard negatives, forcing the model to confront its own failure cases to enhance fine-grained discrimination among highly similar tools. 
To overcome the third challenge, we design a dual semantic boundary-aware dynamic Top-K strategy. By adaptively monitoring reranking score cliffs and inter-tool semantic shifts, this method dynamically truncates the candidate list, maximizing relevant tool recall while filtering long-tail noise. 
Furthermore, since existing benchmarks~\citep{wang2025hammerbench,qin2024toolllm} lack process-level annotations, we construct MTDTool, the first multi-turn dialogue task decomposition benchmark tailored for mobile scenarios, providing a standardized foundation for evaluating decomposition quality.

The contributions of this paper are summarized as follows:
\begin{itemize}
\item \textit{Insight.} We provide a comprehensive diagnosis of the current tool retrieval pipeline for autonomous agents, formulating three critical bottlenecks: the non-causal reward signals in RL-based decomposition, the vulnerability of traditional reranking to functionally similar tools, and the inherent noise-inducing limitations of rigid Top-K truncation.

\item \textit{Method.} We propose MagicSelector, a joint optimization framework that seamlessly integrates task decomposition, fine-grained reranking, and adaptive context truncation. It utilizes counterfactual rewards to inject causal retrieval gains, introduces progressive reranking driven by self-distillation hard negative mining, and develops a dynamic Top-K strategy based on dual semantic boundary awareness.

\item \textit{Dataset.} To bridge the gap in evaluating complex real-world interactions, we construct and release MTDTool. This is the first multi-turn dialogue task decomposition benchmark tailored for mobile scenarios, featuring fine-grained process-level annotations to support holistic pipeline evaluation.

\item \textit{Evaluation.} We conduct extensive evaluations across diverse and realistic tool-use scenarios. Empirical results demonstrate that MagicSelector significantly outperforms existing state-of-the-art approaches in overall retrieval accuracy, exhibiting exceptional robustness and remarkably high token efficiency, particularly in out-of-domain (OOD) tool generalization.
\end{itemize}

\section{Related Work}
\subsection{Planning and Tool Retrieval in Agents}
Research on agent planning and tool retrieval has evolved along two primary paradigms~\citep{huang2024understanding, yehudai2025survey}: (1) Prompt-based and In-Context Learning approaches, grounded in the “Reasoning-Acting-Reflecting” cycle established by ReAct~\citep{yao2022react} and Tool-ReAct~\citep{brauns2025toolreagt}. Subsequent works (e.g., Toolformer~\citep{schick2023toolformer}, Gorilla~\citep{patil2024gorilla}, and the ToolLLM~\citep{qin2024toolllm} series) have achieved effective tool selection within fixed inventories by explicitly generating chains of thought, sub-goals, or API invocation sequences; and (2) Joint optimization via Retrieval-Augmented Generation (RAG)~\citep{zhao2026retrieval} and Reinforcement Learning~\citep{liu2026llm}, designed to address challenges in large-scale open tool libraries. These methods typically decompose complex instructions into structured sub-queries (e.g., DAG-based atomic planning in Plan-RAG~\citep{verma2024plan} or iterative query planning in TOOLQP~\citep{fang2026beyond}) and refine policies using synthetic trajectories and RLVR. Recent studies (e.g., ToolRL~\citep{qian2026toolrl}, RLTR~\citep{li2025encouraging}, GiGPO~\citep{feng2026group}) further incorporate GRPO~\citep{guo2025deepseek}, fine-grained tool-use rewards, or hierarchical advantage estimation to enhance end-to-end performance in long-horizon planning and multi-step invocation.
However, existing paradigms largely rely on sparse rewards based on final task completion or coarse-grained trajectory-level feedback. They often treat task decomposition as a static heuristic preceding retrieval, lacking fine-grained causal modeling of how decomposition semantics directly influence tool retrieval ranking quality. To address this limitation, we model multi-turn contextual task decomposition as an optimizable planning strategy and propose a Counterfactual Retrieval Reward mechanism. By quantifying the relative gain in retrieval quality before and after decomposition, we explicitly inject retrieval ranking signals into policy training, thereby achieving joint collaborative optimization of task planning and tool retrieval.

\subsection{Benchmarks for Tool Retrieval}
Existing benchmarks for agent task decomposition and tool retrieval fall into two primary categories:  tool-interaction benchmarks (e.g., API-Bank~\citep{li2023api}, ToolBench~\citep{qin2024toolllm}, MINT-Bench~\citep{wang2024mint}, $\tau$-Bench~\citep{yao2024tau}, UltraTool~\citep{huang2024planning}, and ToolRet~\citep{shi2025retrieval}), which focus on tool selection, parameter alignment, complex long-tail invocation, and retrieval robustness; and task-oriented benchmarks (e.g., AgentBench~\citep{liu2024agentbench}, GAIA~\citep{mialon2024gaia}, WebArena~\citep{zhou2024webarena}, and TaskBench~\citep{shen2024taskbench}), which emphasize end-to-end planning, task decomposition, and completion rates in open environments. Although recent efforts like HammerBench~\citep{wang2025hammerbench} have extended evaluation to function calls in real-world mobile scenarios, mainstream benchmarks remain constrained by Outcome Bias, relying predominantly on final success rates or trajectory-level metrics. They critically lack fine-grained process annotations for multi-turn interactions, such as cross-turn context inheritance, coreference resolution, dynamic intent evolution, task atomization, and semantic tool mapping. To address this, we introduce MTDTool, a fine-grained benchmark for multi-turn mobile interactions. By providing multi-level annotations covering dialogue intent evolution chains and ranked tool candidates, it establishes a new paradigm for fine-grained process evaluation in complex mobile contexts, offering a standardized basis for quantifying the impact of task decomposition quality on tool retrieval efficacy.

\subsection{LLM-based Reranking}
Large Language Models (LLMs) have significantly advanced text and tool reranking, outperforming traditional encoder-based models. Existing reranking methodologies predominantly fall into three paradigms: pointwise, pairwise, and listwise. \textbf{Pointwise methods} independently score each query-document pair \citep{nogueira2019passage, nogueira2020document, sachan2022}. While highly efficient for establishing a foundational relevance scorer, they inherently ignore the fine-grained interactions among candidate items. \textbf{Pairwise approaches} address this by ranking items through direct pair comparisons \citep{nogueira2019multi, pradeep2021expando, qin2023}, providing stronger ranking signals but incurring prohibitive computational costs. \textbf{Listwise rerankers} jointly evaluate multiple candidates to infer their optimal relative order \citep{sun2023chatgpt, ma2023, Pradeep2023RankVicunaZL}, demonstrating superior performance in capturing holistic list context. However, relying on a single paradigm often struggles to balance efficiency and fine-grained discrimination, especially when distinguishing highly similar tools in out-of-domain scenarios. To bridge this gap, our MagicSelector framework introduces a progressive reranking method that synergizes the strengths of these paradigms. By initially employing point-wise training to build a robust base relevance scorer, and subsequently utilizing list-wise training to optimize the relative order of candidates, MagicSelector effectively captures both absolute and relative relevance. Furthermore, this progressive process is driven by self-distillation hard negative mining, which significantly enhances fine-grained discrimination among highly similar tools.

\subsection{Dynamic Top-K Truncation in Tool Retrieval}
Research on Top-K truncation strategies for tool retrieval primarily explores two paradigms. First, Fixed Top-K Truncation \citep{lewis2020, liu2024lost} relies on a predefined constant limit. This rigid approach causes recall misses when relevant tools exceed this limit, or introduces massive irrelevant noise when they are fewer, wasting context tokens and exacerbating LLM hallucination risks. Second, Threshold-based Dynamic Top-K sets absolute thresholds or relative proportions based on reranking scores \citep{xia2025mmed}. However, score distributions vary drastically across queries, making fixed thresholds prone to over-fitting or under-fitting. This is critical in complex multi-task scenarios where a single user instruction contains multiple independent sub-tasks. Here, true relevant tools for distinct sub-tasks might lack high overall matching scores with the global query. Applying a fixed threshold erroneously filters out these specific target tools, leading to multi-task execution failures.Furthermore, existing dynamic Top-K methods rely on a single strategy \citep{taguchi2025efficient} focusing solely on the absolute relevance between tools and queries. They neglect the relative redundancy and semantic coherence within the retrieved tool list, failing to accurately capture the true semantic cliff between highly relevant tools and irrelevant long-tail noise. To address effective truncation following LLM reranking, we propose a Dual Semantic Boundary-Aware dynamic Top-K strategy that breaks the limitations of fixed truncation quantities and single score thresholds. By jointly perceiving the reranking score cliff, which reflects the decay boundary of absolute relevance, and the semantic cliff between adjacent tools, which reflects the disruption boundary of semantic coherence within the candidate list, we achieve adaptive dynamic Top-K truncation.

\section{Query Decomposition and Rewriting}

\begin{figure}[htbp]
    \centering
\includegraphics[width=1\textwidth]{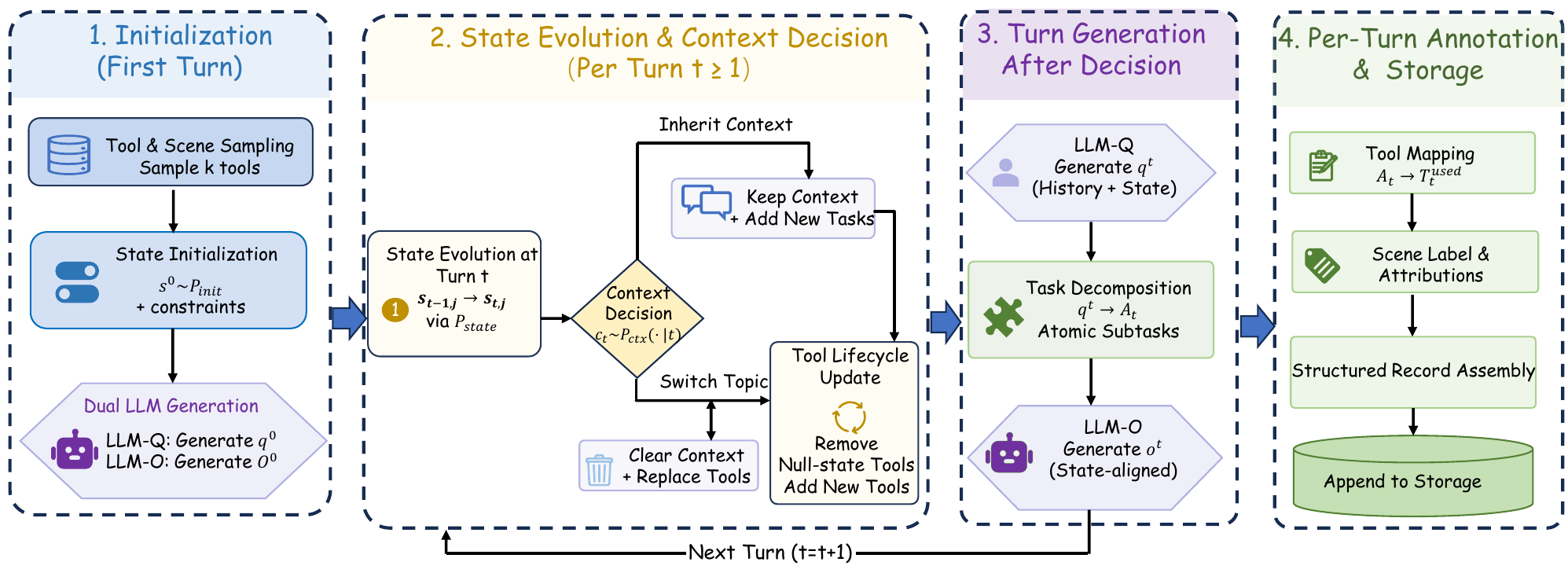}
    \caption{\textbf{Overview of the MTDTool construction pipeline.}
The data generation follows a four-stage process.
\textbf{(1) Initialization}: Tools and scenarios are sampled to establish initial constraints, and two LLM generators produce the initial user query $q^0$ and assistant output $o^0$.
\textbf{(2) State Evolution \& Context Decision}: At each turn $t\geq 1$, the system decides whether to inherit the previous context or switch to a new topic, and updates the pending intent $p_{t,j}$ and execution state $s_{t,j}$ for each active tool.
\textbf{(3) Turn Generation}: Conditioned on the dialogue history $H_t$, current states, and active tools, the framework generates the user query $q_t$, derives the atomic task sequence $\mathcal{A}_t$, and produces the state-aligned assistant output $o_t$.
\textbf{(4) Per-Turn Annotation \& Storage}: The framework maps the atomic task sequence $\mathcal{A}_t$ to the used tool set $\mathcal{T}^{\mathrm{used}}_t$, assigns scenario labels and attribution information, and stores the structured record.
This pipeline preserves the coherence of multi-turn intent evolution while providing fine-grained annotations for task decomposition and tool retrieval.}
    \label{fig:my_picture}
    \vspace{-13pt}
\end{figure}

Query decomposition and rewriting are essential techniques for handling complex user inputs (see Figure~\ref{fig:my_picture}). By breaking down a complicated question into simpler, manageable sub-queries, models can significantly reduce ambiguity. Recent studies show that rewriting user queries using LLMs improves the downstream retrieval accuracy by aligning the query distribution with the document corpus.

\subsection{Dataset Construction via a State-Machine-Driven Framework}

To support the evaluation of multi-turn interaction capabilities for smartphone assistants in complex tool-calling scenarios, this paper proposes an automated dialogue generation framework driven by a state machine. The framework conceptualizes multi-turn dialogue as a stochastic process in which tool execution states evolve over time. By explicitly modeling user intent shifts, context inheritance, and assistant feedback loops, it generates high-fidelity parallel corpora with fine-grained structural annotations. Unlike existing datasets that merely record final task outcomes, our work further provides atomic task sequences and candidate tool mappings, enabling independent diagnosis of planning reasoning and tool retrieval processes.

\subsubsection{State Space Definition and Dynamic Evolution Mechanism}
\label{subsec:state_space}

We define a state space $\mathcal{S}$ of 11 canonical interaction behaviors, encompassing task execution ($s_1, s_2, s_3$), clarification ($s_4, s_5$), and intent management ($s_6, s_7, s_8, s_{10}, s_{11}$). Each tool $t_i \in \mathcal{T}$ is constrained to a semantically compatible subset $\mathcal{S}_{t_i} \subseteq \mathcal{S}$. The initial state is sampled from a prior distribution $P_{\mathrm{init}}$:
\begin{equation}
  s_i^{(0)} \sim P_{\mathrm{init}}(s \mid t_i), \quad s_i^{(0)} \in \mathcal{S}_{t_i}.
\end{equation}
Lightweight structural constraints are applied to ensure multi-turn continuation.

For subsequent turns $r \geq 1$, state evolution follows a two-stage decoupled stochastic process: generating a user intent $d_i^{(r)} \sim P_{\mathrm{todo}}(d \mid s_i^{(r-1)})$, followed by the execution outcome $s_i^{(r)} \sim P_{\mathrm{task}}(s \mid d_i^{(r)})$. This decouples \textit{user intent evolution} from \textit{execution reliability}, enhancing transition controllability and interpretability. Tools collapsing to the null state are removed from the context, enabling dynamic lifecycle management.

\subsubsection{Context Evolution and Task Composition Strategy}
\label{subsec:context_evolution}

To simulate natural topic convergence, we introduce a context decision variable $c^{(r)} \in \{\mathrm{inherit}, \mathrm{switch}\}$ at each turn. The topic-switching probability follows an exponential decay:
\begin{equation}
  P_{\mathrm{switch}}(r) = \max(p_{\min}, p_0 \cdot \alpha^{r-1}),
\end{equation}
where $p_0 = 0.10$, $\alpha = 0.85$, and $p_{\min} = 0.01$. This ensures early-turn diversity while progressively enhancing contextual consistency. 

For task composition, we sample from three structural modes: single-tool, multi-tool collaboration, and multi-intent parallelism. New tasks can be dynamically injected during context inheritance to maintain dialogue density. Tool sampling employs either global random weighting or scene-based knowledge guidance. The latter ensures logical self-consistency of domain-specific tool combinations via predefined or LLM-generated scene templates.

\subsubsection{Dual-Channel Decoupled Generation and Quality Constraints}

Conditioned on state configuration, context, and dialogue history, we employ a dual-channel mechanism to separately synthesize user queries $q^{(r)}$ and assistant responses $o^{(r)}$. The user-side prompt integrates tool lists, state trajectories, style features, and slot-missing indicators, enforcing natural connectors over mechanical delimiters. The assistant-side prompt is strictly governed by state semantics (e.g., confirming success, explaining failures, or remaining silent), ensuring precise alignment with the underlying state machine.

To ensure corpus authenticity, prompts incorporate four hard constraints: (i) ``one-tool-one-request'' atomicity to prevent implicit multi-tasking; (ii) colloquial conversion of tool vocabulary; (iii) temporal expression consistency; and (iv) strict prohibition of redundant expressions (e.g., summarizing or appeasing) during silence states. This quality control layer guarantees both linguistic naturalness and logical rigor.

\subsubsection{Fine-Grained Annotation and Evaluation Support}
\label{subsec:annotation_evaluation}

At the output end of the pipeline, an independent task decomposition module maps user requests containing multiple intents into an atomic task sequence $\mathcal{A}_t$ and establishes semantic associations with the candidate tool set $\mathcal{T}_{\mathrm{cand},t}$:
\begin{equation}
  q_t, H_t \rightarrow \mathcal{A}_t \rightarrow \mathcal{T}_{\mathrm{cand},t}.
\end{equation}
This three-level annotation (\textit{context}--\textit{tasks}--\textit{tools}) supports ranking metrics and precisely localizes model failures (e.g., in context understanding, decomposition, or retrieval), providing a structured diagnostic basis.

The final dataset records comprehensive metadata (e.g., dialogue context, execution states, decomposition results, and tool descriptions), forming a structured corpus for training, evaluation, and attribution analysis. Complete transition matrices, prompt templates, and data formats are detailed in the Appendix.

\begin{figure}[!t]
    \centering
    \includegraphics[width=\textwidth]{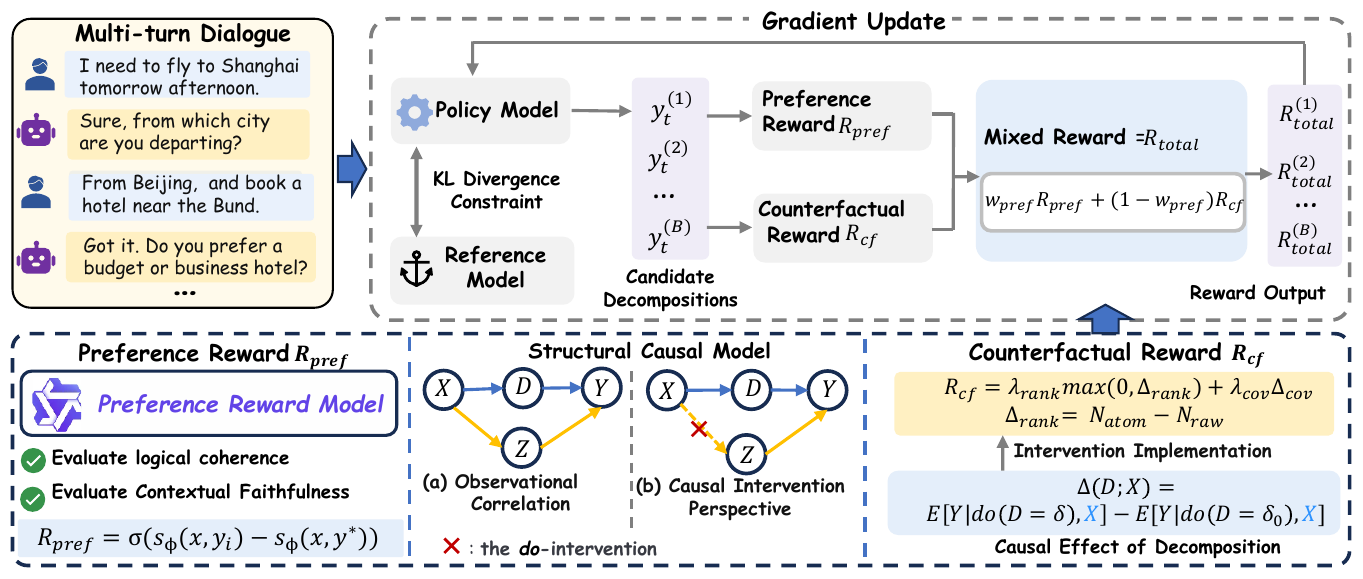}
    \caption{\textbf{Overview of the preference-guided counterfactual task decomposition mechanism in MagicSelector.} It is worth noting that $S_\phi$ denotes the preference reward model, $\mathcal{R}(\cdot)$ represents the downstream tool retrieval model, and the scoring function $\mathcal{M}(\cdot)$ is implemented using NDCG as the evaluation metric.
     }
    \label{fig:algorithm_framework}
    \vspace{-13pt}
\end{figure}

\subsection{Joint Reward Mechanism}
In the scenario of contextual task atomization, the efficacy of model-generated decomposition results should not be determined solely by superficial textual rationality, but rather by their actual contribution to downstream tool retrieval performance and the semantic quality of the decomposition structure itself. To this end, we propose a joint reward mechanism that integrates counterfactual causal feedback with preference alignment, guiding the policy model to learn optimal semantic decomposition paths.




\subsubsection{Counterfactual Baseline Construction and Reward Mechanism}

Given a multi-turn context $x=(H,q)$ and target tools $\mathcal{G}$ (see Figure~\ref{fig:algorithm_framework}), a retriever $\mathcal{R}_K(\cdot)$ outputs a top-$K$ tool list. The non-intervention baseline is $T_{\mathrm{raw}}^K = \mathcal{R}_K(x)$. Its alignment with $\mathcal{G}$ is measured by NDCG@K:
\begin{equation}
 N_{\mathrm{raw}} = \mathrm{NDCG@K}(T_{\mathrm{raw}}^K, \mathcal{G}) = \frac{1}{Z_K} \sum_{j=1}^{K} \frac{2^{\mathbb{I}[t_j \in \mathcal{G}]} - 1}{\log_2(j+1)},
\end{equation}
where $t_j \in T_{\mathrm{raw}}^K$ and $Z_K$ is the ideal normalization factor. $N_{\mathrm{raw}}$ serves as the counterfactual baseline, representing retrieval quality without task decomposition.

For an atomization candidate $y_i \sim \pi_\theta(\cdot|x)$ generated by policy $\pi_\theta$, we parse its sub-tasks $\mathcal{A}(y_i)=\{a_1,\dots,a_m\}$. Retrieving tools independently for each sub-task yields $T_j^K = \mathcal{R}_K(a_j)$. After deduplication and fusion, we obtain the atomized retrieval list $T_{\mathrm{atom}}^K$, with its ranking quality denoted as $N_{\mathrm{atom}} = \mathrm{NDCG@K}(T_{\mathrm{atom}}^K, \mathcal{G})$.

\subsubsection{Positive Counterfactual Gain and Completeness Constraint}
To explicitly quantify the causal impact of task atomization on retrieval efficacy while preserving completeness, we define the counterfactual reward as a linear combination of positive ranking gain and coverage completeness:
$$
R_{\mathrm{cf}} = \lambda_{\mathrm{ndcg}} \max(0, N_{\mathrm{atom}} - N_{\mathrm{raw}}) + \lambda_{\mathrm{full}} (\mathbb{I}[\mathcal{G} \subseteq T_{\mathrm{atom}}^K] - \mathbb{I}[\mathcal{G} \subseteq T_{\mathrm{raw}}^K]).
$$
The first term, $\max(0, N_{\mathrm{atom}} - N_{\mathrm{raw}})$, acts as a unidirectional dense reward. By filtering out non-positive gains, it strictly incentivizes decompositions that outperform the raw query, thereby providing stable gradients and mitigating biases from inherent sample difficulty. The second term utilizes an indicator function to enforce full-coverage completeness, ensuring no essential tools are omitted during ranking optimization. $\lambda_{\mathrm{ndcg}}$ and $\lambda_{\mathrm{full}}$ are weighting hyperparameters.




\subsection{Preference Reward Modeling}
To address the semantic limitations of pure retrieval metrics, we introduce a preference reward $R_{\mathrm{pref}}$ guided by a Process Reward Model (PRM). The PRM evaluates decompositions across five dimensions: completeness, accuracy, coreference resolution, standardized expression, and contextual consistency.

Let $s_\phi(x, y)$ denote the PRM score for a generated result $y$ given context $x$, and $y^\star$ be the human-annotated reference. The preference reward quantifies the relative advantage of the candidate over the reference:
\begin{equation}
R_{\mathrm{pref}} = \sigma\big(s_\phi(x, y_i) - s_\phi(x, y^\star)\big),
\end{equation}
where $\sigma(\cdot)$ is the Sigmoid function. This contrastive design effectively aligns the generated outputs with expert-level logical coherence and instruction adherence.

During GRPO optimization, the objective counterfactual reward and subjective preference reward are fused into the total training signal:
\begin{equation}
R = \alpha R_{\mathrm{pref}} + \beta R_{\mathrm{cf}}.
\end{equation}
This joint formulation ensures the policy model simultaneously optimizes retrieval efficacy and semantic compliance, fostering robust and human-aligned task atomization capabilities.

\subsection{Group Relative Policy Optimization}
To achieve efficient policy optimization without training an independent value network (Critic), we adopt Group Relative Policy Optimization (GRPO)~\citep{guo2025deepseek}. For each input $x_t$, GRPO samples a group of $B$ candidate outputs $\{y_1, \dots, y_B\}$ from the old policy $\pi_{\theta_{old}}$ and computes their corresponding reward scores $\mathbf{R} =\{R_1, \dots, R_B\}$ based on the joint reward formulation defined above. The advantage function for each sample is estimated via group-wise normalization: 
\begin{equation}
A_i = \frac{R_i - \text{mean}(\mathbf{R})}{\text{std}(\mathbf{R}) + \epsilon}.
\end{equation}
This approach leverages relative performance within the group to replace absolute value estimation, significantly reducing memory overhead and enhancing training stability. The optimization objective is defined as:
\begin{equation}
\begin{aligned}
  J_{GRPO}(\theta) = \mathbb{E}_{x \sim \mathcal{D}, \{y_i\} \sim \pi_{\theta_{old}}} \Bigg[ \frac{1}{B} \sum_{i=1}^B \bigg( &\min\left(r_i(\theta)A_i, \text{clip}(r_i(\theta), 1-\epsilon, 1+\epsilon)A_i\right) \\
  &- \beta D_{KL}(\pi_\theta || \pi_{ref}) \bigg) \Bigg],
\end{aligned}
\label{eq: grpo}
\end{equation}
where $r_i(\theta) = \frac{\pi_\theta(y_i \mid x_t)}{\pi_{\theta_{old}}(y_i \mid x_t)}$ is the importance sampling ratio, and $D_{KL}$ denotes the KL divergence penalty relative to a reference model $\pi_{ref}$. In this work, we integrate the proposed joint reward (incorporating the Counterfactual NDCG Reward and Preference Reward) into this framework to drive the targeted evolution of task decomposition strategies.

\section{Retrieval and Reranking}

Practical tool retrieval needs to balance scalability and fine-grained discrimination. Large tool libraries make exhaustive matching impractical, while complex user instructions often differ substantially from tool documentation. To address these difficulties, we adopt a two-stage retrieval and reranking pipeline. The first stage performs embedding-based retrieval over the full tool library, while the second stage applies a cross-encoder reranker to the recalled candidates.

Given a query or decomposed atomic task $q$ and a tool library $\mathcal{T}=\{t_1,\dots,t_N\}$, the embedding retriever first returns a high-recall candidate set:
\begin{equation}
\mathcal{C}_q = \mathcal{R}_{K_r}(q, \mathcal{T}), \qquad \mathcal{C}_q \subset \mathcal{T}.
\end{equation}
This stage reduces the search space from the full library to a smaller set of potentially relevant tools, making subsequent fine-grained matching computationally feasible.

\begin{figure}[htbp]
  \centering
  \includegraphics[width=\textwidth]{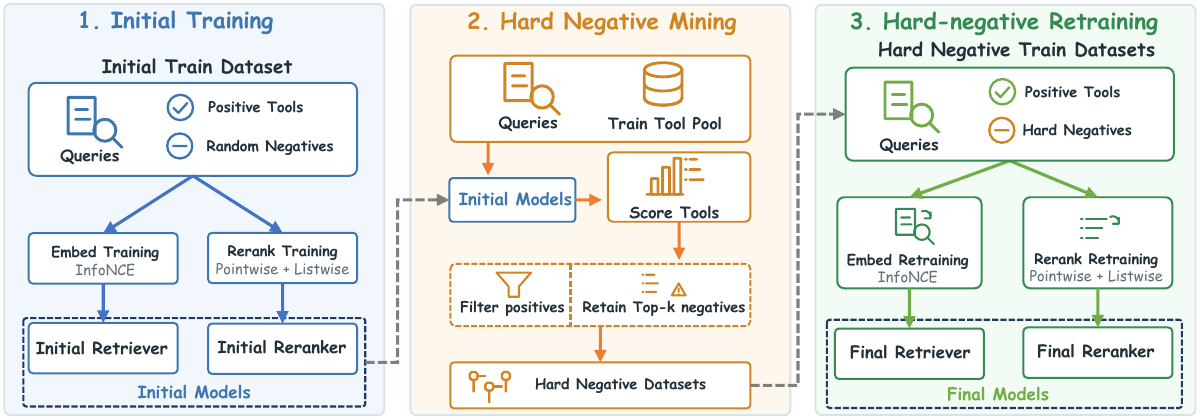}
  \caption{\textbf{Overview of the training workflow of the retrieval and progressive reranking module in MagicSelector.} The module first trains an embedding retriever and a reranker with random negatives, then mines high-scoring incorrect tools from the training tool pool as hard negatives. The hard-negative dataset is used to retrain both models and obtain enhanced retrieval and reranking models.}
  \label{fig:retriever_rerank_workflow}
\end{figure}

The reranker then assigns each candidate tool a relevance score $s(q,t_i)$ and sorts the candidates in descending order:
\begin{equation}
[t_{(1)}, \dots, t_{(m)}] = \mathrm{Top}\text{-}m_{\downarrow}\left(s(q,t_i),\ t_i \in \mathcal{C}_q\right).
\end{equation}
This second stage focuses computation on ambiguous candidates, especially tools with overlapping descriptions or similar functions. During training, self-distillation-driven hard negative mining further selects high-scoring but incorrect tools as informative negatives, improving discrimination without adding online inference cost. Figure~\ref{fig:retriever_rerank_workflow} summarizes the training workflow of the retrieval and reranking module.



\subsection{Embedding-based Retriever Training}

The embedding retriever is designed to handle the scalability requirement of large tool libraries. Since applying a cross-encoder reranker to every tool would be prohibitively expensive, we first encode queries and tool descriptions into a shared dense vector space and perform approximate nearest neighbor search. This stage prioritizes recall rather than final precision. Its goal is to retain potentially relevant tools in the candidate set, including tools whose descriptions differ lexically from the user query but remain semantically related. Training encourages matched query-tool pairs to be closer than irrelevant pairs in this space.

Let $E_\psi(\cdot)$ denote the embedding model parameterized by $\psi$. For a query or decomposed atomic task $q$ and a tool $t$, their dense representations are
\begin{equation}
  \mathbf{e}_q = E_\psi(q), \quad \mathbf{e}_t = E_\psi(t).
\end{equation}
We compute relevance with cosine similarity,
\begin{equation}
  \mathrm{sim}(q,t) = \cos(\mathbf{e}_q, \mathbf{e}_t),
\end{equation}
which is equivalent to inner product when the vectors are normalized.

We train the embedding model with a contrastive objective. Given a query $q$, a positive tool $t^+$, and a set of negative tools $\{t_1^-, \dots, t_M^-\}$, we use the following InfoNCE loss.
\begin{equation}
\mathcal{L}_{\mathrm{emb}}
=
-\log
\frac{
\exp(\mathrm{sim}(q,t^+)/\tau)
}{
\exp(\mathrm{sim}(q,t^+)/\tau)
+
\sum_{j=1}^{M}
\exp(\mathrm{sim}(q,t_j^-)/\tau)
},
\end{equation}
where $\tau$ is a temperature coefficient. The loss pulls matched query-tool pairs closer and pushes irrelevant tools away, producing an embedding space suitable for approximate nearest neighbor search.

During inference, all tool descriptions are encoded offline and indexed by an approximate nearest neighbor search engine. For each query or atomic task, the retriever returns the top-$K_r$ tools by embedding similarity.
\begin{equation}
  \mathcal{C}_q = \mathrm{Top}\text{-}K_r \left( \mathrm{sim}(q,t_i), t_i \in \mathcal{T} \right).
\end{equation}
The candidate set $\mathcal{C}_q$ is then passed to the reranker for more precise relevance estimation.

\subsection{Progressive Reranker Training}

Although the embedding retriever provides efficient recall, its bi-encoder architecture compresses the query and each tool description independently. As a result, it cannot fully model fine-grained token-level interactions between heterogeneous user expressions and tool documentation. This limitation is especially harmful when multiple candidate tools share similar surface descriptions but differ in subtle functional constraints, input schemas, or execution effects. To resolve these ambiguities, we introduce a cross-encoder reranker, which takes each query-tool pair as joint input and outputs a relevance score. Compared with the retriever, the reranker directly models query-tool interactions and is therefore better suited for ordering semantically similar candidates.

Let $R_\omega$ denote the reranker parameterized by $\omega$. Given a query $q$ and a candidate tool $t_i \in \mathcal{C}_q$, the reranker computes
\begin{equation}
  s_i = s(q,t_i) = R_\omega(q,t_i).
\end{equation}
The final tool ranking is obtained by sorting all candidate tools according to $s_i$ in descending order. We train the reranker in two stages. Point-wise training first builds a basic relevance scorer, and list-wise training then optimizes the relative order of tools within the candidate list.



\subsubsection{Point-wise Training}

In the first stage, the reranker learns to judge each query-tool pair independently. Each pair $(q,t_i)$ has a binary relevance label $y_i \in \{0,1\}$, where $y_i=1$ means that $t_i$ is a ground-truth tool for $q$. The point-wise loss is
\begin{equation}
\mathcal{L}_{\mathrm{point}}
=
-
\left[
y_i \log \sigma(s(q,t_i))
+
(1-y_i)\log(1-\sigma(s(q,t_i)))
\right],
\end{equation}
where $\sigma(\cdot)$ denotes the sigmoid function. This stage gives the model a stable relevance scoring function before list-wise ranking optimization.

\subsubsection{List-wise Training}

Point-wise training treats each pair independently, but retrieval quality depends on the relative order of tools in the candidate list. We continue training with a list-wise objective that models the ranking distribution over the whole candidate set.

For a query $q$ and its candidate list $\mathcal{C}_q=\{t_1,\dots,t_n\}$, the reranker produces the score vector
\begin{equation}
  \mathbf{s}_q = [s(q,t_1), \dots, s(q,t_n)].
\end{equation}
We convert the scores into a predicted ranking distribution.
\begin{equation}
  P_i =
  \frac{\exp(s(q,t_i))}
  {\sum_{j=1}^{n}\exp(s(q,t_j))}.
\end{equation}
Given a target distribution $P_i^*$ derived from ground-truth relevance labels, the list-wise loss is
\begin{equation}
  \mathcal{L}_{\mathrm{list}}
  =
  -\sum_{i=1}^{n} P_i^* \log P_i.
\end{equation}
This objective encourages the reranker to place more probability mass on relevant tools and improves the ordering among semantically similar candidates.

\subsection{Self-Distillation Driven Hard Negative Mining}

Negative samples strongly affect both retriever and reranker training. Random negatives are often far from the query and easy for the model to reject, which gives weak learning signals. We use self-distillation-driven hard negative mining to expose the model to harder mistakes. The current model scores candidate tools, and its high-scoring incorrect predictions are selected as hard negatives for the next model. Because the supervision comes from the model itself and the student keeps the same architecture, we call this procedure self-distillation.

The strategy is model-agnostic and can be applied to either the embedding retriever or the reranker. Let $f_\theta$ denote the model being enhanced. For the retriever, $s(q,t)$ is the embedding similarity $\mathrm{sim}(q,t)$. For the reranker, it is the cross-encoder output $R_\omega(q,t)$. We first train an initial model with the corresponding base objective and randomly sampled negative tools. For each query $q$, we construct tuples $(q,t^+,t^-)$, where $t^+$ is a ground-truth tool and $t^-$ is a random negative tool. The induced pairwise loss is
\begin{equation}
  \mathcal{L}_{\mathrm{pair}}
  =
  -\log \sigma(s(q,t^+) - s(q,t^-)).
\end{equation}
This stage yields an initial model $f_{\theta_0}$ with basic semantic matching ability. In practice, the mined negatives can be added to the retriever's contrastive objective or to the reranker's point-wise and list-wise objectives.

After initial training, iteration $r$ uses the current model $f_{\theta_r}$ as a teacher to score candidate tools. To keep teacher scoring tractable, especially for the cross-encoder reranker, we draw the candidate pool $\mathcal{C}_q \subseteq \mathcal{T}$ from the top tools returned by the current retriever instead of scoring the entire library. For each query $q$ and candidate tool $t_i \in \mathcal{C}_q$, the teacher score is
\begin{equation}
  s_r(q,t_i) = f_{\theta_r}(q,t_i).
\end{equation}
The candidate tools are then ranked by these scores.

After removing all ground-truth tools $\mathcal{G}_q$, we select the top-ranked incorrect tools as hard negatives.
\begin{equation}
\mathcal{T}_{\mathrm{hard}}^{-}(q)
=
\mathrm{Top}\text{-}k
\{
t_i
\mid
t_i \in \mathcal{C}_q,\ t_i \notin \mathcal{G}_q,
\text{ ranked by } s_r(q,t_i)
\}.
\end{equation}
These hard negatives receive high relevance scores from the teacher model but are not correct tools. They expose the model's current confusion patterns and provide stronger supervision than random negatives.

The reconstructed training tuples are
\begin{equation}
  (q,t^+,t_{\mathrm{hard}}^-),
  \quad
  t_{\mathrm{hard}}^- \in \mathcal{T}_{\mathrm{hard}}^{-}(q).
\end{equation}
The model is retrained on the new data to obtain an improved student model $f_{\theta_{r+1}}$. The process can be repeated iteratively.
\begin{equation}
  f_{\theta_r}
  \rightarrow
  \text{hard negative mining}
  \rightarrow
  f_{\theta_{r+1}}.
\end{equation}
Through this self-distillation process, the model repeatedly revisits its own failure cases and becomes better at distinguishing semantically similar tools.

\begin{algorithm}[htbp]
\caption{Self-Distillation Driven Hard Negative Mining}
\begin{algorithmic}[1]
\State \textbf{Input} Training queries $\mathcal{Q}$, tool library $\mathcal{T}$, ground-truth tool sets $\{\mathcal{G}_q\}_{q\in\mathcal{Q}}$, hard negative number $k$, iteration number $R$
\State \textbf{Initialize} Construct initial tuples $(q,t^+,t^-)$ with randomly sampled negatives $t^- \notin \mathcal{G}_q$
\State Train an initial model $f_{\theta_0}$ using the corresponding base training objective
\For{$r = 0$ \textbf{to} $R-1$}
    \State \textbf{Teacher scoring}
    \For{each query $q \in \mathcal{Q}$}
        \State Retrieve a candidate pool $\mathcal{C}_q \subseteq \mathcal{T}$ with the current retriever
        \State Compute teacher scores $s_r(q,t_i)=f_{\theta_r}(q,t_i)$ for each candidate tool $t_i \in \mathcal{C}_q$
        \State Rank candidate tools according to $s_r(q,t_i)$ in descending order
        \State Remove all ground-truth tools $\mathcal{G}_q$ from the ranked list
        \State Select the top-$k$ remaining tools as hard negatives
        \Statex \hspace{1.5em} $\mathcal{T}^{-}_{\mathrm{hard}}(q)=\mathrm{Top}\text{-}k\{t_i \mid t_i \in \mathcal{C}_q,\ t_i \notin \mathcal{G}_q, \text{ ranked by } s_r(q,t_i)\}$
    \EndFor
    \State \textbf{Training data reconstruction}
    \State Construct hard-negative training tuples $(q,t^+,t^-_{\mathrm{hard}})$, where $t^+ \in \mathcal{G}_q$ and $t^-_{\mathrm{hard}} \in \mathcal{T}^{-}_{\mathrm{hard}}(q)$
    \State \textbf{Student retraining}
    \State Train a new student model $f_{\theta_{r+1}}$ with the reconstructed hard-negative training tuples
\EndFor
\State \textbf{Return} Enhanced retriever or reranker $f_{\theta_R}$
\end{algorithmic}
\end{algorithm}

For the retriever, hard negatives improve the dense retrieval space by forcing the model to separate tools with similar surface descriptions but different functions. For the reranker, they strengthen fine-grained relevance discrimination among highly similar candidate tools. The process does not require extra human annotation, so it scales naturally to large tool libraries.

Overall, the module has two coupled but distinct flows. Training starts from initial retriever and reranker models, mines hard negatives through self-distillation, and produces enhanced models. Inference applies the enhanced retriever and reranker in a simple cascade.
\begin{equation}
\text{High-recall Retrieval}
\rightarrow
\text{High-precision Reranking}
\rightarrow
\text{Ranked Tool List}.
\end{equation}
This design keeps online retrieval efficient while improving the retriever's recall structure and the reranker's precision.

\section{Adaptive Top-k}
In tool retrieval tasks for Large Language Models (LLMs), traditional fixed Top-K truncation methods struggle to balance recall and contextual redundancy across queries of varying complexity. To address this issue, this paper proposes a dual semantic boundary-aware dynamic Top-K algorithm. This algorithm treats the reranked tool list as a one-dimensional semantic space and adaptively determines the optimal truncation position by jointly evaluating the absolute relevance decay between the query and tools, as well as the relative semantic gap between adjacent tools.

Specifically, our proposed truncation strategy based on the tool-tool relative semantic gap is inspired by the TextTiling algorithm~\citep{hearst1997text}. We generalize the classical mechanism of detecting similarity gaps between adjacent text blocks into the tool semantic space constructed by dense retrieval models. By integrating the tool functional clustering characteristics revealed by ToolLLM~\citep{ICLR2024_28e50ee5} and Anytool~\citep{pmlr-v235-du24h}, alongside the notion from Rankvicuna~\citep{Pradeep2023RankVicunaZL} which conceptualizes the reranked list as a one-dimensional sequence with local contextual coherence, our approach demonstrates that the embedding similarity gap between functional descriptions of adjacent tools in the reranked list serves as an effective signal for identifying tool functional boundaries.

\subsection{Characteristics of Tool Reranking Score Distribution Based on the Pareto Distribution}

In the retrieval and reranking tasks for large-scale tool libraries, we observe that the distribution of tool reranking scores exhibits a significant long-tail characteristic, closely approximating a Pareto distribution~\citep{2005Power}. Specifically, for a given complex user query, only a minuscule fraction of tools within the tool semantic space can precisely match the user intent, forming the high-scoring head; conversely, the vast majority of tools are functionally orthogonal or exhibit only weak marginal relevance to the query, causing their scores to decay rapidly and constitute a massive low-scoring long tail.

\begin{figure}[htbp]
  \centering
  \includegraphics[width=0.6\textwidth]{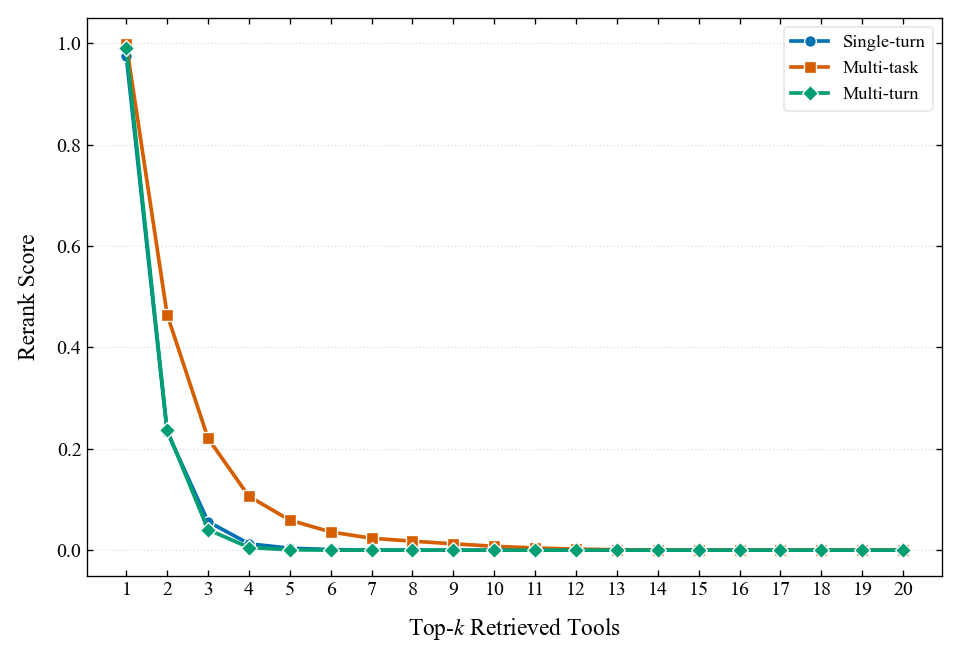}
  \caption{Reranking score distribution of retrieved tools}
  \label{fig:main_agent_rerank2}
\end{figure}

As illustrated in Figure~\ref{fig:main_agent_rerank2}, we conduct an empirical analysis of the reranking score distribution for tools recalled by the agent across different scenarios. The statistics encompass a single-turn dataset (6,930 instances), a multi-task dataset (2,990 instances), and a multi-turn dataset (3,706 instances). The observations reveal that, regardless of whether the scenario involves simple single-turn queries or complex multi-task and multi-turn interactions, the reranking scores consistently demonstrate a highly uniform Pareto-like long-tail decay. This phenomenon intuitively reflects that only a select few core tools attain high matching scores in real-world interactive tasks, whereas the overwhelming majority of candidate tools rapidly decay and cluster in the low-scoring tail region.

Let the reranking score be a random variable $X$. Under the assumption of a Pareto distribution, its probability density function can be approximately expressed as:
\begin{equation}
f(x) = \frac{\alpha x_m^\alpha}{x^{\alpha+1}} \quad (x \ge x_m)
\end{equation}
where $x_m$ is the lower bound threshold of the relevance score, and $\alpha$ is the shape parameter determining the decay rate of the tail. This distributional characteristic profoundly reveals the extreme imbalance of relevance distribution within the tool semantic space.

The steep decay characteristic of the Pareto distribution in the head region provides robust theoretical support for our proposed dynamic Top-K strategy based on dual semantic boundary awareness. Since the actual relevance scores often experience a sharp, precipitous drop after a few head tools, the traditional fixed-$K$ truncation method ignores this non-uniform distribution and is highly prone to crossing the inflection point into the long-tail region, thereby introducing substantial redundant noise into the context of the large language model.

In contrast, leveraging the natural gap characteristic of the Pareto distribution, our algorithm jointly evaluates the absolute relevance decay and the relative semantic gap between adjacent tools. This enables the model to adaptively strip away long-tail noise, maximizing the semantic purity of the input context while ensuring the recall rate of core tools.

\subsection{Recall-Maximizing Fusion Strategy}

Given a user query $q$ and an ordered list of candidate tools $T = [t_1, \dots, t_n]$ output by a reranking model. For any tool $t_i$ in the list, its relevance score with respect to the query is denoted as $s_i = \text{score}(q, t_i)$. Since $T$ is sorted in descending order based on $s_i$, this list inherently forms a one-dimensional relevance space.

\begin{algorithm}[htbp]
\caption{Recall-Maximizing Fusion Strategy}
\label{alg:fusion}
\begin{algorithmic}[1]
\State \textbf{Input:} Sorted candidate tools $T = \{t_1, \dots, t_n\}$, reranking scores $S = \{s_1, \dots, s_n\}$, tool description embeddings $E = \{e_1, \dots, e_n\}$
\State \textbf{Initialize:} $K_{\text{score}} \leftarrow 0$, $K_{\text{sim}} \leftarrow 0$
\State \textbf{Score-based Truncation:}
\State Calculate score gaps $\Delta s_i = s_i - s_{i+1}$ for $i \in \{1, \dots, n-1\}$
\State $K_{\text{score}} = \arg\max_i (\Delta s_i)$ \# Truncate at the maximum reranking score gap
\State \textbf{Semantic-based Truncation:}
\State Calculate adjacent tool similarities $sim_i = \text{CosineSim}(e_i, e_{i+1})$ for $i \in \{1, \dots, n-1\}$
\State Calculate similarity gaps $\Delta sim_i = sim_i - sim_{i+1}$ for $i \in \{1, \dots, n-2\}$ \# Measure the semantic drop between adjacent tools
\State $K_{\text{sim}} = \arg\max_i (\Delta sim_i)$ \# Truncate at the maximum semantic similarity gap
\State \textbf{Recall-Maximizing Fusion:}
\State $K_{\text{final}} = \max(K_{\text{score}}, K_{\text{sim}})$ \# Take the maximum $K$ to ensure high recall
\State \textbf{Return:} Final truncated tool list $T_{\text{final}} = \{t_1, \dots, t_{K_{\text{final}}}\}$
\end{algorithmic}
\end{algorithm}

To capture the degradation of relevance within the tool list, we compute the first-order difference of the scores between adjacent tools. The position exhibiting the maximum score gap typically signifies the boundary separating relevant tools from noisy ones~\citep{taguchi2025efficient,xia2025mmed}. Therefore, the candidate truncation point based on the score difference is defined as:
\begin{equation}
\mathrm{Top\_K}_{\mathrm{score}} = \underset{i}{\arg\max} (s_i - s_{i+1})
\end{equation}

Relying solely on the degradation of $s_i$ is prone to failure when the scores decrease smoothly. To address this, we further conceptualize the reranked list as a one-dimensional continuous semantic space and introduce a metric for the semantic coherence between adjacent tools. 

Let $sim_i = \text{sim}(t_i, t_{i+1})$ denote the cosine similarity between the functional description embeddings of adjacent tools $t_i$ and $t_{i+1}$. Within this one-dimensional sequence, a sharp drop in $sim_i$ indicates an abrupt shift in tool or functional clusters. We interpret such abrupt shifts as clustering boundaries within the one-dimensional space, and compute the point of maximum similarity difference as the second candidate truncation point:
\begin{equation}
\mathrm{Top\_K}_{\mathrm{sim}} = \underset{i}{\arg\max} (\mathrm{sim}_i - \mathrm{sim}_{i+1})
\end{equation}

Considering the low tolerance for false negatives in tool retrieval tasks, we adopt a conservative boundary fusion strategy. We define the final truncation position as the maximum of the two aforementioned candidate truncation points:
\begin{equation}
\mathrm{Final\_Top\_K} = \max(\mathrm{Top\_K}_{\mathrm{score}}, \mathrm{Top\_K}_{\mathrm{sim}})
\end{equation}
When the degradation of relevance scores is not pronounced, $\mathrm{Top\_K}_{\mathrm{sim}}$ can acutely capture the semantic discontinuities in the underlying tool functions. Conversely, when local tool functions are similar but the overall relevance has already dropped significantly, $\mathrm{Top\_K}_{\mathrm{score}}$ provides a fundamental truncation safeguard. The operation of taking the maximum essentially maximizes the recall of potentially relevant tools under the premise of ensuring that semantic coherence and relevance remain intact.The complete procedure of our proposed method is summarized in Algorithm~\ref{alg:fusion}.

Furthermore, by only computing the score and semantic differences between adjacent tools, the time complexity of our strategy is strictly bounded to $O(n)$, avoiding the high computational overhead of $O(n^2)$ associated with global pairwise comparisons.


\section{Experiments}


We comprehensively evaluate the proposed framework across a diverse suite of benchmarks that capture complementary dimensions of tool-oriented agents: ToolBench for realistic API invocations, ToolRet for large-scale tool retrieval, and MTDTool for multi-turn mobile interactions. Our empirical study is specifically designed to rigorously assess the efficacy of MagicSelector in retrieving relevant tools from extensive repositories, optimizing candidate ranking, and demonstrating robust generalization across varied query formulations and annotation paradigms.


\subsection{Evaluation Benchmarks}

\textbf{ToolBench}~\citep{qin2024toolllm}. As a comprehensive real-world API benchmark, ToolBench features 16,464 RESTful APIs across 49 categories and roughly 126k instruction-solution pairs spanning diverse single- and multi-tool scenarios. In our evaluation, it serves as a rigorous testbed to assess whether the agent can accurately retrieve prerequisite tools from a massive structured library, a crucial step that dictates downstream execution performance.

\textbf{ToolRet.} ToolRet~\citep{shi2025retrieval} aggregates 35 datasets across Web, Code, and Custom domains, featuring a corpus of 44K tools. Following ToolQP, its training set comprises 10K Web-domain instances with trajectories synthesized via Qwen3.6-Plus. For evaluation, 3.1K Web samples serve as the In-Domain test set (tools exposed during training), while the remainder forms the Out-Domain test set.

\textbf{MTDTool.} MTDTool specifically constructed for complex multi-turn mobile interactions, encompassing 237 vertical-domain tools. Its test set is correspondingly partitioned into In-Domain and out-of-domain subsets based on tool vertical domains.


\subsection{Baselines}
To comprehensively evaluate task decomposition and tool retrieval performance, we mainly compare against four categories of baselines: (1) closed-source LLMs (GPT-5.0\footnote{https://chat.openai.com/}, MiniMax-M2.5\footnote{https://minimaxi.com/news/minimax-m25}, and DeepSeek-V4-Pro\footnote{https://www.deepseek.com/}) to establish zero-shot and few-shot performance upper bounds; (2) prompting-based methods (Q2E~\citep{wang2023query2doc}, ReInvoke~\citep{chen2024re}, ToolReAGt~\citep{brauns2025toolreagt}, and PLUTO~\citep{huang2024pluto}), representing training-free paradigms that leverage structured retrieval augmentation or chain-of-thought; (3) ToolQP~\citep{fang2026beyond}, an RL-based method that jointly optimizes decomposition and retrieval via mixed reward signals; and (4) dense retrievers and rerankers (gte-Qwen2-7B-instruct~\citep{gteqwen2_7b_modelcard,li2023gte}, E5-Mistral-7B-Instruct~\citep{wang2024improving_text}, and their enhanced versions with Qwen3-Reranker-8B), which serve as strong retrieval-only baselines to evaluate the fundamental search capability.


\subsection{Experimental Configuration}


By default, our retrieval and reranking backbone employs Qwen3-Embedding-4B for dense retrieval and Qwen3-Reranker-4B for cross-encoder reranking~\citep{qwen3embedding}.

\textbf{ToolBench configuration.} Following ToolOmni~\citep{huang2026toolomni}, we evaluate ToolBench under three instruction difficulty levels (I1--I3) and three generalization splits: Instruction, Tool, and Category. We consider two settings: \textit{in-domain}, where models are trained and tested within the same domain, and \textit{multi-domain}, where models are trained on the full dataset. Unless otherwise specified, retrieval is conducted in an open-domain setting, where each query retrieves tools from the complete tool corpus rather than a pre-filtered API list. Baselines include BM25~\citep{robertson2009bm25}, EmbSim~\citep{reimers2019sentencebert}, Re-Invoke~\citep{chen2024reinvoke}, IterFeedback~\citep{xu2024iterfeedback}, ToolGen~\citep{wang2025toolgen}, ToolRetriever~\citep{qin2024toolllm}, ToolOmni~\citep{huang2026toolomni}, gte-Qwen2-7B-instruct, and E5-Mistral-7B-Instruct. Retrieval performance is measured by NDCG@$k$, where $k \in \{1,3,5\}$.

\textbf{ToolRet configuration.} Following the setting of Tool-DE~\citep{lu2025tools}, both training and evaluation use the instruction-augmented ToolRet format. All methods retrieve from the full tool library, and we report NDCG@10 (N@10), Recall@10 (R@10), and Completeness@10 (C@10).

\textbf{MTDTool configuration.} MTDTool is our self-constructed domain-specific dataset for multi-turn mobile tool retrieval. We evaluate in-domain and out-of-domain splits defined by tool vertical domains. The comparison includes four selected embedding--reranking combinations: Qwen3-Embedding-8B with Qwen3-Reranker-8B~\citep{qwen3embedding}, gte-Qwen2-7B-instruct with Qwen3-Reranker-8B, E5-Mistral-7B-Instruct with Qwen3-Reranker-8B, and Qwen3-Embedding-4B with Qwen3-Reranker-4B. We report NDCG@10, Recall@10, and Completeness@10.

\begin{table}[!ht]
\centering
\small
\setlength{\tabcolsep}{2pt}
\renewcommand{\arraystretch}{1.1}
\caption{Overall performance comparison of MagicSelector and baselines across benchmarks. Best results are bolded.}
\label{tab:overall_performance}
\begin{tabular}{lccccccc}
\toprule
\multirow{2}{*}{\textbf{Methods}}
& \multicolumn{2}{c}{\textbf{ToolRet}}
& \multicolumn{2}{c}{\textbf{MTDTool}}
& \multicolumn{3}{c}{\textbf{ToolBench}} \\

\cmidrule(lr){2-3}
\cmidrule(lr){4-5}
\cmidrule(lr){6-8}

& N@10 & C@10
& N@10 & C@10
& N@1 & N@3 & N@5 \\

\midrule
\multicolumn{8}{l}{\textit{\textbf{Base Retriever}}} \\
Qwen3-Embedding-4B & 45.54 & 47.27 & 63.59 & 54.87 & 21.2 & 33.3 & 38.0 \\
\midrule
\multicolumn{5}{l}{\textit{\textbf{Closed-Source LLMs}}} \\
MiniMax-M2.5 & 43.73 & 45.46 & 79.31 & 74.68 & 30.8 & 32.2 & 34.9 \\
DeepSeek-V4-Pro & 45.48 & 47.65 & 78.16 & 73.96 & 31.5 & 33.6 & 36.8 \\
ChatGPT-5.0 & 47.25 & 49.98 & 78.02 & 78.37 & 33.2 & 34.4 & 36.9 \\
\midrule
\multicolumn{8}{l}{\textit{\textbf{Prompting Methods (Qwen3-4B)}}} \\
Q2E & 39.61 & 40.90 & 68.37 & 59.59 & 43.7 & 40.8 & 44.3 \\
ReInvoke & 44.92 & 46.81 & 82.00 & 76.26 & 39.1 & 33.2 & 37.4 \\
ToolReAGt & 38.52 & 43.78 & 74.82 & 66.70 & 42.0 & 42.2 & 46.2 \\
PLUTO & 42.74 & 43.18 & 77.49 & 57.20 & 53.8 & 47.9 & 50.7 \\
\midrule
\multicolumn{5}{l}{\textit{\textbf{RL-Based Methods (Qwen3-4B)}}} \\
ToolQP & 45.36 & 49.30 & 80.66 & 72.64 & 47.6 & 44.8 & 47.7 \\
\midrule
\multicolumn{8}{l}{\textit{\textbf{Prompting Methods (Qwen3-8B)}}} \\
Q2E & 44.44 & 46.06 & 64.81 & 57.90 & 44.4 & 40.7 & 44.2 \\
ReInvoke & 43.96 & 45.88 & 77.46 & 73.76 & 40.7 & 35.3 & 38.4 \\
ToolReAGt & 40.31 & 44.74 & 79.09 & 72.35 & 44.4 & 43.4 & 46.9 \\
PLUTO & 40.52 & 42.96 & 76.42 & 61.05 & 53.5 & 47.2 & 49.7 \\
\midrule
\multicolumn{5}{l}{\textit{\textbf{RL-Based Methods (Qwen3-8B)}}} \\
ToolQP & 47.12 & 51.30 & 83.82 & 78.97 & 45.4 & 42.8 & 46.6 \\
\midrule
\multicolumn{8}{l}{\textit{\textbf{Dense Retrievers \& Rerankers}}} \\
gte-Qwen2-7B-instruct & 40.93 & 41.59 & 62.54 & 56.44 & 14.0 & 18.0 & 18.9 \\
\quad + Qwen3-Reranker-8B & 51.04 & 50.03 & 78.88 & 83.15 & 20.5 & 22.2 & 22.4 \\
E5-Mistral-7B-Instruct & 39.85 & 40.57 & 62.45 & 56.16 & 47.3 & 55.3 & 54.5 \\
\quad + Qwen3-Reranker-8B & 51.39 & 50.08 & 79.53 & 83.66 & 52.8 & 59.6 & 59.6 \\
\rowcolor{blue!8} Ours-Embedding-4B & 54.23 & 53.71 & 91.36 & 87.65 & 83.1 & 87.8 & 87.1 \\
\rowcolor{blue!8} \quad + Ours-Rerank-4B & 59.44 & 58.90 & 96.00 & 95.71 & \textbf{85.7} & \textbf{90.7} & 90.6 \\
\rowcolor{blue!8} \quad + Ours-Decomp + Ours-Rerank-4B & \textbf{59.90} & \textbf{59.21} & \textbf{96.28} & \textbf{96.01} & 84.2 & \textbf{90.7} & \textbf{90.8} \\

\bottomrule
\end{tabular}
\end{table}






\begin{table}[!ht]
\centering
\small
\renewcommand{\arraystretch}{1.1}
\caption{Overall performance comparison on the MTDTool benchmark under in-domain and out-of-domain settings. Best and second-best results are bolded and underlined, respectively.}
\label{tab:mtdtool_merged_results}
\begin{tabularx}{\textwidth}{l *{4}{>{\centering\arraybackslash}X}}
\toprule
\multirow{2}{*}{\textbf{Methods}} & \multicolumn{2}{c}{\textbf{In-Domain}} & \multicolumn{2}{c}{\textbf{Out-of-Domain}} \\
\cmidrule(lr){2-3}
\cmidrule(lr){4-5}
 & \textbf{N@10} & \textbf{C@10} & \textbf{N@10} & \textbf{C@10} \\
\midrule
\multicolumn{5}{l}{\textit{\textbf{Base Retriever}}} \\
Qwen3-Embedding-4B & 69.49 & 62.10 & 57.69 & 47.64 \\
\midrule
\multicolumn{5}{l}{\textit{\textbf{Closed-Source LLMs}}} \\
MiniMax-M2.5 & 84.11 & 79.18 & 74.51 & 70.18 \\
DeepSeek-V4-Pro & 83.06 & 78.70 & 73.26 & 69.22 \\
ChatGPT-5.0 & 82.81 & 82.50 & 73.22 & 74.24 \\
\midrule
\multicolumn{5}{l}{\textit{\textbf{Prompting Methods (Qwen3-4B)}}} \\
Q2E & 74.97 & 67.50 & 61.77 & 51.68 \\
ReInvoke & 84.27 & 77.87 & 79.73 & 74.64 \\
ToolReAGt & 77.09 & 68.81 & 72.54 & 64.58 \\
PLUTO & 77.69 & 56.29 & 77.29 & 58.11 \\
\midrule
\multicolumn{5}{l}{\textit{\textbf{RL-Based Methods (Qwen3-4B)}}} \\
ToolQP & 86.72 & 76.99 & 74.60 & 68.28 \\
\rowcolor{blue!8} Ours-Decomp & 89.17 & 84.11 & 78.82 & 74.87 \\
\midrule
\multicolumn{5}{l}{\textit{\textbf{Prompting Methods (Qwen3-8B)}}} \\
Q2E & 73.01 & 66.16 & 56.61 & 49.64 \\
ReInvoke & 80.27 & 76.21 & 74.65 & 71.31 \\
ToolReAGt & 81.20 & 74.57 & 76.98 & 70.12 \\
PLUTO & 76.91 & 61.73 & 75.93 & 60.36 \\
\midrule
\multicolumn{5}{l}{\textit{\textbf{RL-Based Methods (Qwen3-8B)}}} \\
ToolQP & 89.72 & 83.61 & 77.92 & 74.33 \\
\rowcolor{blue!8} Ours-Decomp & 91.19 & 87.08 & 82.74 & 79.77 \\
\midrule
\multicolumn{5}{l}{\textit{\textbf{Dense Retrievers \& Rerankers}}} \\
gte-Qwen2-7B-instruct & 65.89 & 59.62 & 59.18 & 53.26 \\
\quad + Qwen3-Reranker-8B & 81.30 & 83.78 & 76.46 & 82.52 \\
E5-Mistral-7B-Instruct & 69.42 & 64.97 & 55.47 & 47.35 \\
\quad + Qwen3-Reranker-8B & 81.84 & 85.08 & 77.21 & 82.23 \\
Qwen3-Embedding-8B & 71.27 & 63.43 & 63.99 & 55.21 \\
\quad + Qwen3-Reranker-8B & 81.82 & 84.90 & 78.40 & 85.29 \\
\rowcolor{blue!8} \quad + Ours-Rerank-4B & 97.01 & 93.20 & 94.88 & 93.44 \\
Qwen3-Embedding-4B & 69.49 & 62.10 & 57.69 & 47.64 \\
\quad + Qwen3-Reranker-4B & 82.39 & 85.33 & 77.52 & 81.48 \\
\rowcolor{blue!8} \quad + Ours-Rerank-4B & 96.88 & 92.51 & 93.52 & 89.72 \\
\rowcolor{blue!8} \textbf{Ours} & \textbf{97.22} & \textbf{95.33} & \textbf{95.33} & \textbf{96.69} \\
\bottomrule
\end{tabularx}
\end{table}

\subsection{Main Results}

We report the main retrieval results on ToolBench, ToolRet, and MTDTool in Tables~\ref{tab:overall_performance}--\ref{tab:toolret_results}. These benchmarks cover complementary retrieval scenarios: ToolBench evaluates API-level tool selection under realistic tool-use instructions, ToolRet measures large-scale retrieval over heterogeneous tool domains, and MTDTool focuses on multi-turn mobile interactions with in-domain and out-of-domain splits. Unless otherwise specified, methods retrieve from the full tool corpus defined by each benchmark or split. Across these benchmarks, we compare three model configurations: an embedding-only retriever, an embedding--reranking pipeline, and a full pipeline that further incorporates the decomposition model before retrieval.

\begin{table}[!ht]
\centering
\footnotesize 
\setlength{\tabcolsep}{6pt} 
\renewcommand{\arraystretch}{1.1} 
\caption{ToolBench retrieval performance (In-Domain) measured by NDCG@$k$ (\%).}
\label{tab:toolbench_ndcg_indomain}
\begin{tabularx}{\textwidth}{l *{9}{>{\centering\arraybackslash}X} c}
\toprule
\multirow{2}{*}{\textbf{Methods}} & \multicolumn{3}{c}{\textbf{I1}} & \multicolumn{3}{c}{\textbf{I2}} & \multicolumn{3}{c}{\textbf{I3}} & \multirow{2}{*}{\textbf{Avg.}} \\
\cmidrule(lr){2-4} \cmidrule(lr){5-7} \cmidrule(lr){8-10}
 & \textbf{N@1} & \textbf{N@3} & \textbf{N@5} & \textbf{N@1} & \textbf{N@3} & \textbf{N@5} & \textbf{N@1} & \textbf{N@3} & \textbf{N@5} & \\
\midrule
\multicolumn{11}{l}{\textit{\textbf{Base Retriever}}} \\
Qwen3-Embedding-4B & 64.2 & 80.0 & 82.7 & 48.5 & 68.1 & 73.0 & 50.0 & 68.5 & 73.3 & 67.6 \\
\midrule
\multicolumn{11}{l}{\textit{\textbf{Other Retriever Methods}}} \\
BM25 & 29.5 & 31.1 & 33.3 & 24.1 & 25.3 & 27.7 & 32.0 & 25.9 & 29.8 & 28.7 \\
EmbSim & 63.7 & 61.0 & 65.4 & 49.1 & 42.3 & 46.6 & 53.0 & 46.4 & 52.7 & 53.4 \\
ToolGen & 69.5 & 72.3 & 79.1 & 46.8 & 53.6 & 62.5 & 77.1 & 76.4 & 85.5 & 69.2 \\
IterFeedback & 71.6 & 71.3 & 76.3 & 62.7 & 55.6 & 60.6 & 73.9 & 64.1 & 69.0 & 67.2 \\
ToolRetriever & 81.9 & 82.1 & 85.6 & 75.9 & 69.6 & 75.4 & 79.8 & 72.4 & 77.1 & 77.8 \\
ToolOmni & 86.1 & 84.5 & 85.6 & 81.5 & 73.1 & 74.1 & 84.4 & 76.8 & 77.3 & 80.4 \\
\midrule
\multicolumn{11}{l}{\textit{\textbf{Closed-Source LLMs}}} \\
MiniMax-M2.5 & 81.1 & 81.6 & 84.8 & 85.0 & 78.4 & 83.4 & 83.0 & 78.2 & 84.7 & 82.2 \\
DeepSeek-V4-Pro & 83.5 & 82.9 & 86.6 & 85.0 & 77.2 & 83.5 & 85.0 & 80.9 & 86.5 & 83.5 \\
ChatGPT-5.0 & 84.1 & 82.5 & 86.2 & 86.5 & 78.9 & 84.2 & 86.0 & 81.4 & 87.4 & 84.1 \\
\midrule
\multicolumn{11}{l}{\textit{\textbf{Prompting Methods (Qwen3-4B)}}} \\
Q2E & 81.1 & 77.1 & 81.0 & 61.5 & 48.2 & 48.8 & 46.0 & 38.4 & 39.0 & 57.9 \\
ReInvoke & 72.3 & 67.5 & 68.4 & 64.0 & 51.3 & 51.2 & 56.0 & 47.4 & 46.1 & 58.2 \\
ToolReAGt & 83.1 & 79.3 & 80.6 & 63.0 & 48.4 & 48.8 & 63.0 & 48.1 & 47.7 & 62.4 \\
PLUTO & 87.2 & 78.8 & 80.1 & 67.5 & 51.4 & 51.2 & 70.0 & 53.5 & 51.6 & 65.7 \\
\midrule
\multicolumn{11}{l}{\textit{\textbf{RL-Based Methods (Qwen3-4B)}}} \\
ToolQP & 86.1 & 84.5 & 89.6 & 78.1 & 74.5 & 85.0 & 71.3 & 63.2 & 76.1 & 78.7 \\
\midrule
\multicolumn{11}{l}{\textit{\textbf{Prompting Methods (Qwen3-8B)}}} \\
Q2E & 80.4 & 76.8 & 80.5 & 63.5 & 48.2 & 48.8 & 47.0 & 40.7 & 40.6 & 58.5 \\
ReInvoke & 72.3 & 67.2 & 67.7 & 64.0 & 51.2 & 51.1 & 63.0 & 49.3 & 48.3 & 59.3 \\
ToolReAGt & 83.1 & 81.5 & 83.2 & 66.0 & 50.6 & 50.4 & 67.0 & 49.6 & 48.4 & 64.4 \\
PLUTO & 85.8 & 77.7 & 78.9 & 68.0 & 49.5 & 49.5 & 74.0 & 55.2 & 53.5 & 65.8 \\
\midrule
\multicolumn{11}{l}{\textit{\textbf{RL-Based Methods (Qwen3-8B)}}} \\
ToolQP & 86.1 & 84.8 & 88.9 & 76.6 & 74.3 & 84.5 & 61.4 & 61.0 & 74.0 & 76.9 \\
\midrule
\multicolumn{5}{l}{\textit{\textbf{Dense Retrievers \& Rerankers}}} \\
gte-Qwen2-7B-instruct & 76.0 & 83.8 & 85.3 & 72.5 & 83.6 & 85.4 & 55.0 & 75.2 & 78.8 & 77.3 \\
\quad + Qwen3-Reranker-8B & 90.2 & 94.5 & 93.2 & 89.5 & 93.6 & 89.0 & 85.0 & 92.2 & 91.0 & 90.9 \\
E5-Mistral-7B-Instruct & 87.8 & 94.2 & 93.6 & 85.0 & 92.5 & 90.9 & 81.0 & 90.1 & 88.6 & 89.3 \\
\quad + Qwen3-Reranker-8B & 92.2 & 96.1 & 95.7 & 91.0 & 95.7 & 93.9 & 89.0 & 94.0 & 92.2 & 93.3 \\
\rowcolor{blue!8} \textbf{Ours} & \textbf{98.0} & \textbf{98.8} & \textbf{98.2} & \textbf{96.0} & \textbf{97.1} & \textbf{96.5} & \textbf{94.0} & \textbf{95.9} & \textbf{94.8} & \textbf{96.6} \\
\bottomrule
\end{tabularx}
\end{table}

As illustrated in Table~\ref{tab:overall_performance}, we report the overall retrieval performance of our proposed MagicSelector framework against various baselines across three diverse benchmarks: ToolRet, MTDTool, and ToolBench. The results represent average values across specific subsets for each benchmark. Specifically, the MTDTool metrics are the average of in-domain and out-of-domain splits, the ToolBench metrics represent the average of multi-domain I1, I2, and I3 scenarios, and the ToolRet metrics are the average of web, code, and customized domains. Detailed results are presented in Tables~\ref{tab:mtdtool_merged_results}--\ref{tab:toolret_results}. MagicSelector consistently achieves state-of-the-art performance, demonstrating superior efficacy over both prompting-based methods and specialized dense retrievers.

Our foundational embedding model establishes a remarkably strong baseline that significantly outperforms all prompting-based methods such as ReInvoke, Q2E, ToolReAGt, and PLUTO, as well as competitive dense retrievers like gte-Qwen2 and E5-Mistral. For instance, on ToolRet, it achieves 54.23 N@10 and 53.71 C@10, which substantially surpasses the best prompting method ReInvoke at 44.92 N@10 and the reranked E5-Mistral baseline at 51.39 N@10. Similar massive advantages are observed on MTDTool with 91.36 N@10 compared to 82.00 N@10, and on ToolBench with 83.1 N@1 compared to 53.8 N@1.

Furthermore, integrating our specialized reranking module provides exceptional refinement capabilities. This addition boosts the N@10 on MTDTool from 91.36 to 96.00, and significantly improves performance across ToolBench to achieve the highest N@1 score of 85.7.

The complete pipeline, which synergistically combines our query decomposition model with the specialized reranker, yields the highest overall metrics. It sets a new state-of-the-art across the datasets, peaking at 59.90 N@10 and 59.21 C@10 on ToolRet, and 96.28 N@10 and 96.01 C@10 on MTDTool. On ToolBench, the decomposition strategy provides crucial additional gains for broader candidate pools, reaching top results of 90.7 N@3 and 90.8 N@5. This validates that combining structural query decomposition with precise reranking provides the most robust and adaptable architecture for complex tool retrieval.


\begin{table}[!ht]
\centering
\footnotesize 
\setlength{\tabcolsep}{6pt} 
\renewcommand{\arraystretch}{1.1} 
\caption{ToolBench retrieval performance (Multi-Domain) measured by NDCG@$k$ (\%).}
\label{tab:toolbench_ndcg_multidomain}
\begin{tabularx}{\textwidth}{l *{9}{>{\centering\arraybackslash}X} c}
\toprule
\multirow{2}{*}{\textbf{Methods}} & \multicolumn{3}{c}{\textbf{I1}} & \multicolumn{3}{c}{\textbf{I2}} & \multicolumn{3}{c}{\textbf{I3}} & \multirow{2}{*}{\textbf{Avg.}} \\
\cmidrule(lr){2-4} \cmidrule(lr){5-7} \cmidrule(lr){8-10}
 & \textbf{N@1} & \textbf{N@3} & \textbf{N@5} & \textbf{N@1} & \textbf{N@3} & \textbf{N@5} & \textbf{N@1} & \textbf{N@3} & \textbf{N@5} & \\
\midrule
\multicolumn{11}{l}{\textit{\textbf{Base Retriever}}} \\
Qwen3-Embedding-4B & 27.0 & 41.2 & 46.7 & 17.5 & 28.4 & 32.8 & 19.0 & 30.2 & 34.6 & 30.8 \\
\midrule
\multicolumn{11}{l}{\textit{\textbf{Other Retriever Methods}}} \\
BM25 & 22.8 & 22.6 & 25.6 & 18.3 & 20.7 & 22.2 & 10.0 & 10.1 & 12.3 & 18.3 \\
EmbSim & 54.0 & 50.8 & 55.9 & 40.8 & 36.7 & 39.6 & 18.0 & 17.8 & 20.7 & 37.1 \\
ToolGen & 68.8 & 72.0 & 78.8 & 46.8 & 53.6 & 62.4 & 75.7 & 75.3 & 84.5 & 68.6 \\
IterFeedback & 71.8 & 70.5 & 74.8 & 63.2 & 55.4 & 61.3 & 67.4 & 59.2 & 63.1 & 65.2 \\
ToolRetriever & 81.0 & 80.9 & 84.5 & 75.9 & 69.5 & 75.2 & 77.5 & 69.2 & 74.2 & 76.4 \\
ToolOmni & \textbf{86.1} & 84.1 & 84.8 & 80.6 & 73.0 & 73.9 & 78.9 & 71.4 & 71.8 & 78.3 \\
\midrule
\multicolumn{11}{l}{\textit{\textbf{Closed-Source LLMs}}} \\
MiniMax-M2.5 & 37.5 & 37.7 & 40.9 & 27.0 & 30.2 & 32.8 & 28.0 & 28.7 & 31.1 & 32.7 \\
DeepSeek-V4-Pro & 38.5 & 39.8 & 42.6 & 28.0 & 32.1 & 34.3 & 28.0 & 28.9 & 33.5 & 34.0 \\
ChatGPT-5.0 & 38.2 & 40.4 & 42.9 & 30.5 & 32.3 & 34.9 & 31.0 & 30.4 & 32.9 & 34.8 \\
\midrule
\multicolumn{11}{l}{\textit{\textbf{Prompting Methods (Qwen3-4B)}}} \\
Q2E & 70.6 & 67.8 & 73.4 & 35.5 & 31.2 & 34.4 & 25.0 & 23.4 & 25.2 & 42.9 \\
ReInvoke & 52.4 & 42.9 & 48.0 & 38.0 & 32.3 & 36.0 & 27.0 & 24.5 & 28.3 & 36.6 \\
ToolReAGt & 57.4 & 59.8 & 65.4 & 34.5 & 34.1 & 37.5 & 34.0 & 32.7 & 35.6 & 43.4 \\
PLUTO & 68.9 & 64.3 & 69.0 & 46.5 & 40.4 & 42.3 & 46.0 & 38.9 & 40.9 & 50.8 \\
\midrule
\multicolumn{11}{l}{\textit{\textbf{RL-Based Methods (Qwen3-4B)}}} \\
ToolQP & 71.2 & 70.2 & 74.9 & 37.2 & 34.0 & 36.6 & 34.3 & 30.3 & 31.6 & 46.7 \\
\midrule
\multicolumn{11}{l}{\textit{\textbf{Prompting Methods (Qwen3-8B)}}} \\
Q2E & 70.6 & 67.7 & 72.5 & 39.5 & 32.0 & 35.4 & 23.0 & 22.3 & 24.8 & 43.1 \\
ReInvoke & 50.7 & 44.4 & 48.6 & 38.5 & 32.1 & 35.0 & 33.0 & 29.5 & 31.5 & 38.1 \\
ToolReAGt & 56.8 & 60.8 & 67.2 & 34.5 & 33.1 & 35.9 & 42.0 & 36.4 & 37.5 & 44.9 \\
PLUTO & 62.5 & 61.5 & 65.5 & 46.0 & 38.6 & 40.7 & 52.0 & 41.4 & 43.0 & 50.1 \\
\midrule
\multicolumn{11}{l}{\textit{\textbf{RL-Based Methods (Qwen3-8B)}}} \\
ToolQP & 69.6 & 69.1 & 74.1 & 35.2 & 31.7 & 35.8 & 31.3 & 27.7 & 29.9 & 44.9 \\
\midrule
\multicolumn{5}{l}{\textit{\textbf{Dense Retrievers \& Rerankers}}} \\
gte-Qwen2-7B-instruct & 17.9 & 23.0 & 23.3 & 18.0 & 21.5 & 22.2 & 6.0 & 9.6 & 11.3 & 17.0 \\
\quad + Qwen3-Reranker-8B & 26.0 & 27.7 & 28.1 & 21.5 & 24.1 & 24.3 & 14.0 & 14.9 & 14.9 & 21.7 \\
E5-Mistral-7B-Instruct & 54.2 & 62.2 & 61.5 & 46.8 & 56.9 & 55.9 & 41.0 & 46.9 & 46.0 & 52.4 \\
\quad + Qwen3-Reranker-8B & 61.8 & 68.4 & 68.3 & 50.5 & 59.5 & 59.9 & 46.0 & 50.8 & 50.7 & 57.3 \\
\rowcolor{blue!8} \textbf{Ours} & \textbf{86.1} & \textbf{91.8} & \textbf{92.1} & \textbf{87.0} & \textbf{91.5} & \textbf{91.4} & \textbf{79.4} & \textbf{88.7} & \textbf{89.0} & \textbf{88.5} \\
\bottomrule
\end{tabularx}
\end{table}

\begin{table}[!ht]
\centering
\caption{Retrieval performance on ToolRet under the instruction-based setting.}
\label{tab:toolret_results}
\tiny
\setlength{\tabcolsep}{1.5pt}
\renewcommand{\arraystretch}{0.9}
\resizebox{\textwidth}{!}{
\begin{tabular}{lcccccccccccc}
\toprule
\multirow{2}{*}{\textbf{Methods}} 
& \multicolumn{3}{c}{\textbf{Web}} 
& \multicolumn{3}{c}{\textbf{Code}} 
& \multicolumn{3}{c}{\textbf{Customized}} 
& \multicolumn{3}{c}{\textbf{Avg.}} \\
\cmidrule(lr){2-4}
\cmidrule(lr){5-7}
\cmidrule(lr){8-10}
\cmidrule(lr){11-13}
 & \textbf{N@10} & \textbf{R@10} & \textbf{C@10} & \textbf{N@10} & \textbf{R@10} & \textbf{C@10} & \textbf{N@10} & \textbf{R@10} & \textbf{C@10} & \textbf{N@10} & \textbf{R@10} & \textbf{C@10} \\
\midrule
\multicolumn{5}{l}{\textit{\textbf{Base Retriever}}} \\
Qwen3-Embedding-4B & 40.90 & 50.14 & 31.97 & 53.56 & 71.12 & 69.61 & 42.15 & 50.80 & 40.22 & 45.54 & 57.36 & 47.27 \\
\midrule
\multicolumn{5}{l}{\textit{\textbf{Other Retriever Methods}}} \\
BM25 & 26.33 & 34.22 & 22.79 & 41.90 & 56.49 & 55.39 & 41.16 & 48.60 & 38.90 & 36.46 & 46.44 & 39.03 \\
COLT & 28.91 & 40.64 & 38.83 & 20.06 & 27.78 & 18.84 & 31.29 & 42.19 & 34.01 & 26.75 & 36.87 & 30.56 \\
Colbert & 16.67 & 21.14 & 14.94 & 30.35 & 41.38 & 40.28 & 24.35 & 30.97 & 24.87 & 23.79 & 31.16 & 26.70 \\
contriever-msmarco & 23.48 & 30.21 & 19.69 & 31.61 & 43.01 & 41.74 & 21.93 & 27.28 & 23.04 & 25.67 & 33.50 & 28.16 \\
gtr-t5-base & 20.38 & 27.53 & 19.24 & 33.59 & 43.18 & 41.88 & 41.84 & 48.35 & 39.28 & 31.94 & 39.69 & 33.46 \\
gtr-t5-large & 24.37 & 31.64 & 21.26 & 36.76 & 47.42 & 45.92 & 42.04 & 50.84 & 40.00 & 34.39 & 43.30 & 35.73 \\
all-MiniLM-L6-v2 & 12.77 & 19.38 & 13.33 & 31.59 & 43.86 & 42.25 & 32.24 & 43.55 & 32.34 & 25.53 & 35.60 & 29.31 \\
E5-small-v2 & 26.42 & 34.44 & 21.39 & 32.36 & 42.38 & 41.11 & 34.62 & 42.29 & 32.58 & 31.14 & 39.70 & 31.69 \\
E5-base-v2 & 24.71 & 33.45 & 21.94 & 31.40 & 42.83 & 41.38 & 38.06 & 46.84 & 36.43 & 31.39 & 41.04 & 33.25 \\
E5-large-v2 & 23.62 & 32.19 & 21.80 & 34.27 & 44.42 & 43.19 & 43.32 & 52.30 & 41.42 & 33.73 & 42.97 & 35.47 \\
gte-base-en-v1.5 & 30.75 & 39.44 & 25.88 & 41.68 & 53.96 & 51.64 & 37.95 & 46.57 & 38.10 & 36.79 & 46.66 & 38.54 \\
gte-large-en-v1.5 & 28.06 & 36.32 & 22.57 & 35.77 & 49.56 & 47.71 & 37.27 & 47.98 & 35.84 & 33.70 & 44.62 & 35.37 \\
bge-base-en-v1.5 & 25.95 & 35.12 & 23.40 & 35.15 & 45.74 & 44.32 & 43.20 & 53.54 & 42.29 & 34.77 & 44.80 & 36.67 \\
bge-large-en-v1.5 & 30.03 & 39.28 & 25.63 & 41.53 & 52.76 & 51.18 & 43.90 & 51.79 & 42.24 & 38.49 & 47.94 & 39.68 \\
E5-Mistral-7b & 31.07 & 41.30 & 27.04 & 44.97 & 58.95 & 56.79 & 40.88 & 49.35 & 38.35 & 38.97 & 49.87 & 40.73 \\
NV-Embed-v1 & 31.51 & 40.52 & 26.74 & 47.92 & 62.07 & 59.60 & 48.70 & 57.69 & 43.88 & 42.71 & 53.43 & 43.41 \\
gte-Qwen2-1.5B-inst. & 37.53 & 48.31 & 30.95 & 47.38 & 61.12 & 59.55 & 52.98 & 59.47 & 45.68 & 45.96 & 56.30 & 45.39 \\
GritLM-7B & 36.58 & 46.01 & 27.65 & 41.26 & 53.81 & 52.07 & 45.55 & 54.01 & 41.40 & 41.13 & 51.28 & 40.37 \\
mxbai-rerank-large-v1 & 17.53 & 25.82 & 17.95 & 33.86 & 47.84 & 46.47 & 26.83 & 37.61 & 28.60 & 26.08 & 37.09 & 31.01 \\
monoT5-base-msmarco & 23.33 & 30.70 & 18.13 & 31.39 & 45.18 & 42.53 & 37.77 & 46.63 & 39.70 & 28.72 & 40.84 & 33.45 \\
bge-rerank-v2-m3 & 34.83 & 45.23 & 31.73 & 50.86 & 67.26 & 64.78 & 42.35 & 53.75 & 39.90 & 42.68 & 55.41 & 45.47 \\
jina-rerank-v2-base & 42.35 & 51.21 & 34.23 & 53.21 & 66.03 & 63.94 & 45.94 & 57.96 & 45.41 & 47.17 & 58.40 & 47.86 \\
bge-rerank-v2-gemma & 34.73 & 45.08 & 32.29 & 55.85 & 70.53 & 68.76 & 51.97 & 61.20 & 45.65 & 47.52 & 58.94 & 48.90 \\
Mixtral-8x22B & 35.31 & 38.63 & 34.60 & 33.27 & 39.60 & 38.53 & 34.40 & 39.72 & 38.20 & 34.33 & 39.32 & 37.11 \\

\midrule
\multicolumn{13}{l}{\textit{\textbf{Closed-Source LLMs}}} \\

MiniMax-M2.5 & 50.18 & 62.21 & 49.71 & 29.53 & 41.58 & 39.68 & 37.84 & 46.13 & 35.23 & 43.73 & 55.28 & 45.46 \\
DeepSeek-v4-pro & 49.16 & 61.12 & 49.09 & 37.92 & 51.63 & 49.40 & 41.15 & 49.58 & 37.58 & 45.48 & 57.38 & 47.65 \\
GPT-5.0 & 50.72 & 63.10 & 50.97 & 42.67 & 58.23 & 54.60 & 38.64 & 48.27 & 36.97 & 47.25 & 60.01 & 49.98 \\

\midrule
\multicolumn{13}{l}{\textit{\textbf{Prompting Methods (Qwen3-4B)}}} \\
Q2E & 43.71 & 54.24 & 42.72 & 34.42 & 45.49 & 42.37 & 29.04 & 37.02 & 29.53 & 39.61 & 49.94 & 40.90 \\
ReInvoke & 49.57 & 61.61 & 49.54 & 35.33 & 49.42 & 45.63 & 39.50 & 47.16 & 35.74 & 44.92 & 56.86 & 46.81 \\
ToolReAGt & 41.97 & 56.16 & 44.63 & 32.68 & 50.14 & 47.34 & 32.19 & 42.47 & 33.30 & 38.52 & 52.96 & 43.78 \\
PLUTO & 46.40 & 56.31 & 43.73 & 40.51 & 51.80 & 49.40 & 29.01 & 36.57 & 29.43 & 42.74 & 52.66 & 43.18 \\

\midrule
\multicolumn{13}{l}{\textit{\textbf{RL-Based Methods (Qwen3-4B)}}} \\
ToolQP & 47.30 & 61.14 & 48.99 & 45.51 & 61.19 & 58.03 & 35.72 & 45.44 & 35.23 & 45.36 & 59.09 & 49.30 \\

\midrule
\multicolumn{13}{l}{\textit{\textbf{Prompting Methods (Qwen3-8B)}}} \\
Q2E & 46.89 & 58.08 & 45.47 & 42.52 & 56.98 & 54.03 & 36.04 & 44.79 & 34.73 & 44.44 & 56.08 & 46.06 \\
ReInvoke & 47.72 & 60.09 & 47.94 & 36.83 & 50.27 & 46.43 & 38.46 & 46.42 & 34.93 & 43.96 & 56.00 & 45.88 \\
ToolReAGt & 43.46 & 57.16 & 45.71 & 35.16 & 50.09 & 47.68 & 34.22 & 45.06 & 34.83 & 40.31 & 53.92 & 44.74 \\
PLUTO & 42.57 & 54.70 & 42.40 & 42.57 & 54.52 & 51.92 & 26.98 & 36.27 & 29.74 & 40.52 & 52.24 & 42.96 \\

\midrule
\multicolumn{13}{l}{\textit{\textbf{RL-Based Methods (Qwen3-8B)}}} \\
ToolQP & 47.81 & 61.61 & 49.56 & 44.89 & 61.34 & 56.26 & 35.56 & 45.28 & 35.34 & 45.52 & 59.41 & 49.26 \\
\midrule
\multicolumn{5}{l}{\textit{\textbf{Dense Retrievers \& Rerankers}}} \\
gte-Qwen2-7B-instruct & 37.74 & 47.86 & 30.62 & 46.20 & 60.46 & 58.82 & 38.85 & 45.82 & 35.34 & 40.93 & 51.38 & 41.59 \\
\quad + Qwen3-Reranker-8B & 43.75 & 54.63 & 38.24 & 58.18 & 70.04 & 67.98 & 51.19 & 56.10 & 43.87 & 51.04 & 60.26 & 50.03 \\
E5-Mistral-7B-Instruct & 32.35 & 42.39 & 27.63 & 44.12 & 57.97 & 55.63 & 43.08 & 50.14 & 38.46 & 39.85 & 50.17 & 40.57 \\
\quad + Qwen3-Reranker-8B & 42.06 & 51.08 & 35.74 & 56.41 & 67.91 & 65.15 & 55.71 & 63.23 & 49.35 & 51.39 & 60.74 & 50.08 \\
Tool-DE-Embed-4B & 45.60 & 55.50 & 37.62 & 54.66 & 69.51 & 67.87 & 54.11 & 63.96 & 48.53 & 51.46 & 62.99 & 51.34 \\
\quad + Tool-DE-Rerank-4B & 48.37 & 59.72 & 43.78 & 59.93 & 75.88 & 74.38 & \textbf{62.72} & \textbf{68.92} & \textbf{52.79} & 57.01 & 68.17 & 56.98 \\
\rowcolor{blue!8} \textbf{Ours} & \textbf{52.78} & \textbf{61.52} & \textbf{45.32} & \textbf{66.63} & \textbf{81.58} & \textbf{79.95} & 60.30 & 67.20 & 52.36 & \textbf{59.90} & \textbf{70.10} & \textbf{59.21} \\
\bottomrule
\end{tabular}
}
\end{table}

Table~\ref{tab:mtdtool_merged_results} reports results on MTDTool, which evaluates multi-turn mobile tool retrieval under both in-domain and out-of-domain settings. For the off-the-shelf embedding backbones, adding a Qwen3 reranker consistently improves N@10 and C@10, highlighting the importance of cross-encoder reranking for distinguishing tools with similar descriptions and schemas. For example, adding Qwen3-Reranker-8B to Qwen3-Embedding-8B improves N@10 from 71.27 to 81.82 and C@10 from 63.43 to 84.90 in the in-domain split. Our RL-based decomposition model Ours-Decomp provides a stronger starting point than standard prompting methods, achieving 91.19 N@10 and 87.08 C@10 in the in-domain split, and 82.74 N@10 and 79.77 C@10 in the out-of-domain split. Combining off-the-shelf embeddings with our specialized reranker Ours-Rerank-4B further improves completeness, especially on the out-of-domain split where C@10 reaches 93.44. After integrating both components, the full pipeline Ours achieves the best in-domain results with 97.22 N@10 and 95.33 C@10, and the best out-of-domain performance with 95.33 N@10 and 96.69 C@10, suggesting that decomposition can further improve ranking when user instructions involve multi-turn or compound intents.

On ToolBench, Tables~\ref{tab:toolbench_ndcg_indomain} and \ref{tab:toolbench_ndcg_multidomain} show that our proposed approach significantly improves performance under both in-domain and multi-domain evaluations. In the in-domain setting, our method achieves 96.6 average NDCG, substantially outperforming the 93.3 average NDCG of the E5-Mistral-7B-Instruct and Qwen3-Reranker-8B baseline. Our approach achieves the highest scores across all I1, I2, and I3 metrics, such as 98.0 NDCG@1 for I1 and 96.0 NDCG@1 for I2. In the multi-domain setting, our framework similarly improves the average score over prior baselines, reaching 88.5 average NDCG compared to 78.3 for ToolOmni, and achieving the best overall results with notable gains in I1 and I3, including 92.1 NDCG@5 for I1 and 89.0 NDCG@5 for I3. These results indicate that our proposed method establishes a strong retrieval foundation and effectively improves candidate API ordering against a broader and more complex tool pool.

Table~\ref{tab:toolret_results} evaluates retrieval performance on the diverse ToolRet library. Our proposed method achieves the best overall results, surpassing the strongest baseline pipeline with average scores of 59.90 N@10, 70.10 R@10, and 59.21 C@10. Specifically, our approach significantly outperforms all other models in the Web domain with 52.78 N@10 and the Code domain with 66.63 N@10. Although slightly trailing the best baseline in the Customized domain, our method secures the highest overall efficacy, demonstrating a robust and adaptable retrieval foundation for complex tool environments.

Overall, the results support three observations. First, a trained embedding retriever provides a strong high-recall foundation for tool retrieval. Second, reranking generally improves the final ordering of candidate tools, with clear gains in ranking and completeness metrics. Third, the decomposition model often brings additional ranking improvements in settings involving complex or multi-turn instructions, suggesting that query decomposition and tool reranking are complementary components rather than interchangeable alternatives.

\begin{figure}[htbp]
 \centering
 \includegraphics[width=0.6\textwidth]{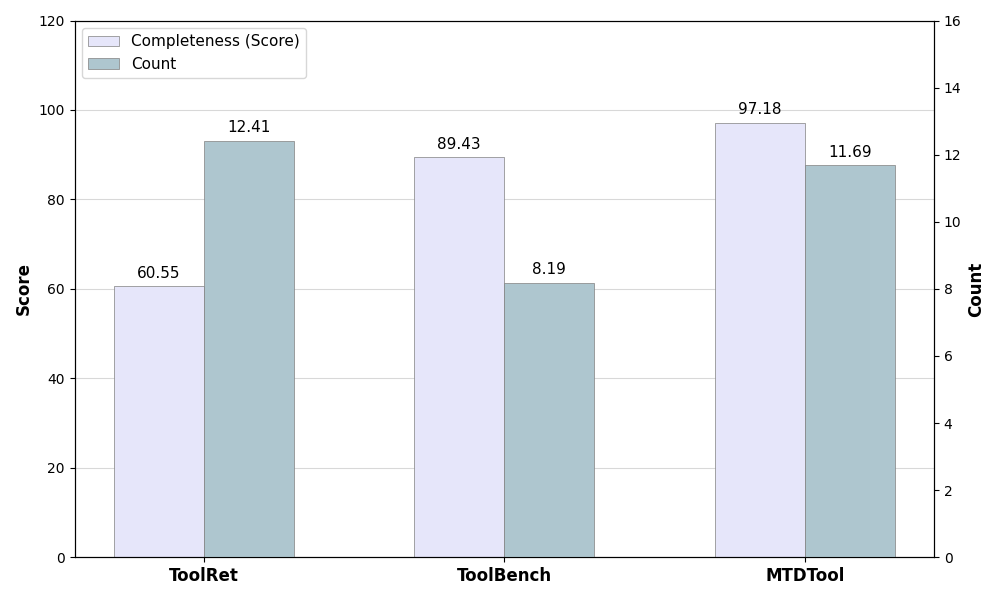}
 \caption{Performance comparison with dynamic Top-K across datasets.}
 \label{fig:all_dynamic_topk}
\end{figure}

Table~\ref{tab:dynamic_topk_datasets} and Figure~\ref{fig:all_dynamic_topk} present the performance comparison between the traditional Completeness@10 (fixed-10) and the dynamic Completeness@k (adaptive-k) settings across three different datasets: ToolRet, ToolBench, and MTDTool. A consistent trend across all evaluated benchmarks is that the adaptive-k strategy consistently outperforms the fixed-10 approach. Specifically, compared to the standard Completeness@10, the adaptive-k setting yields performance improvements of 1.34, 2.13, and 1.17 points on ToolRet, ToolBench, and MTDTool, respectively.

\begin{table}[!ht]
\centering
\small
\setlength{\tabcolsep}{6pt}
\renewcommand{\arraystretch}{1.1}
\caption{Performance comparison with dynamic Top-K across datasets.}
\label{tab:dynamic_topk_datasets}
\begin{tabular}{lcc}
\toprule
\multirow{2}{*}{\textbf{Dataset}} & \multicolumn{2}{c}{\textbf{Avg}} \\
\cmidrule(lr){2-3}
& \textbf{fixed-10} & \textbf{adaptive-k} \\
\midrule
ToolRet & 59.21 & 60.55 \\
ToolBench & 87.30 & 89.43 \\
MTDTool & 96.01 & 97.18 \\
\bottomrule
\end{tabular}
\end{table}

Furthermore, the bar chart in Figure~\ref{fig:all_dynamic_topk} explicitly illustrates how the system adaptively adjusts the average number of retrieved tools  to achieve these scores. For example, the system achieves a high Completeness score of 89.43 on ToolBench while maintaining the lowest average tool count of 8.19. In contrast, for datasets requiring more comprehensive toolsets, it intelligently expands the retrieval size to 12.41 on ToolRet and 11.69 on MTDTool, successfully pushing the Completeness score on MTDTool to an impressive 97.18. These results demonstrate that dynamically adjusting the number of retrieved tools (Completeness@k) is more effective and flexible than using a rigid cutoff (Completeness@10), allowing the retrieval system to better accommodate queries of varying complexity and thereby enhancing the overall retrieval quality across different datasets.

\subsection{Dataset Evaluation}


We categorize MTDTool into two primary dialogue types, single-turn and multi-turn, which are further subdivided into 10 fine-grained scenario labels. Evaluation results in Table~\ref{tab:dataset_quality} based on MagicSelector demonstrate that the model achieves an average accuracy of 89.05\% on the in-domain test set, significantly outperforming the 78.04\% observed on the out-of-domain set. This performance gap validates the effectiveness of our dataset construction and its capacity to sufficiently elicit model capabilities.

Specifically, regarding tool invocation, performance exhibits a gradient decline consistent with increasing complexity. The Single Tool scenario yields the highest performance with an N@10 of 95.48. Although the Cross-category Multi-tool scenario presents greater difficulty, the model maintains an In-Domain N@10 of 86.02 and an OOD N@10 of 74.41, reflecting the dataset's quality and discriminative power in covering concurrent multi-intent and cross-domain orchestration tasks. In terms of multi-turn dialogue, the User Adds New Task and Domain Switch scenarios achieve high accuracy, with In-Domain N@10 exceeding 95, confirming that the dataset effectively supports training for topic switching and new intent recognition. Furthermore, the State Hybrid scenario serves as a high-difficulty benchmark, recording an In-Domain N@10 of 70.35 and an OOD N@10 of 53.36. Similarly, the Unfinished Task Continuation and Abnormal Scenario categories maintain stable OOD performance within the 64--67 range. These outcomes indicate that the dataset successfully encompasses complex logic involving long-term memory, state reasoning, and non-standard execution flows, thereby providing a high-quality and challenging evaluation baseline. In conclusion, the differentiated performance of MagicSelector across various scenarios fully corroborates the high-quality construction of the MTDTool dataset in both breadth and depth.

\begin{table}[htbp]
\centering
\small
\setlength{\tabcolsep}{4pt}
\renewcommand{\arraystretch}{1.3}
\caption{Dataset Quality Evaluation Results}
\label{tab:dataset_quality}
\begin{tabular}{clcccc}
\hline
\multirow{2}{*}{\textbf{Scenario}} & \multirow{2}{*}{\textbf{Type}} & \multicolumn{2}{c}{In-Domain} & \multicolumn{2}{c}{OOD-Domain} \\
\cmidrule(lr){3-4}
\cmidrule(lr){5-6}
 &  & N@10 & C@10 & N@10 & C@10 \\
\hline

\multirow{3}{*}{Single-turn}
& Single Tool & 95.48 & 100.00 & 81.95 & 95.36 \\
& Same-category Multi-tool & 93.99 & 95.42 & 77.77 & 63.78 \\
& Cross-category Multi-tool & 86.02 & 73.27 & 74.41 & 55.48 \\
\hline
\multirow{7}{*}{Multi-turn}
& Domain Switch & 91.78 & 95.89 & 81.27 & 93.08 \\
& Multi-turn Continuation & 81.29 & 83.83 & 69.31 & 74.79 \\
& Intent Selection & 95.09 & 99.29 & 82.88 & 94.13 \\
& User Adds New Task & 94.31 & 98.58 & 84.64 & 89.51 \\
& Unfinished Task Continuation & 90.71 & 87.28 & 73.00 & 64.64 \\
& Abnormal Scenario & 90.20 & 92.68 & 77.89 & 81.65 \\
& State Hybrid & 88.48 & 70.35 & 76.79 & 53.36 \\

\hline
- & Avg & 89.05 & 85.24 & 78.04 & 75.37 \\
\hline
\end{tabular}
\end{table}


\subsection{Deeper Analysis}

\begin{figure*}[htbp]
\centering
\begin{subfigure}{0.24\textwidth}
    \centering
    \includegraphics[width=\linewidth]{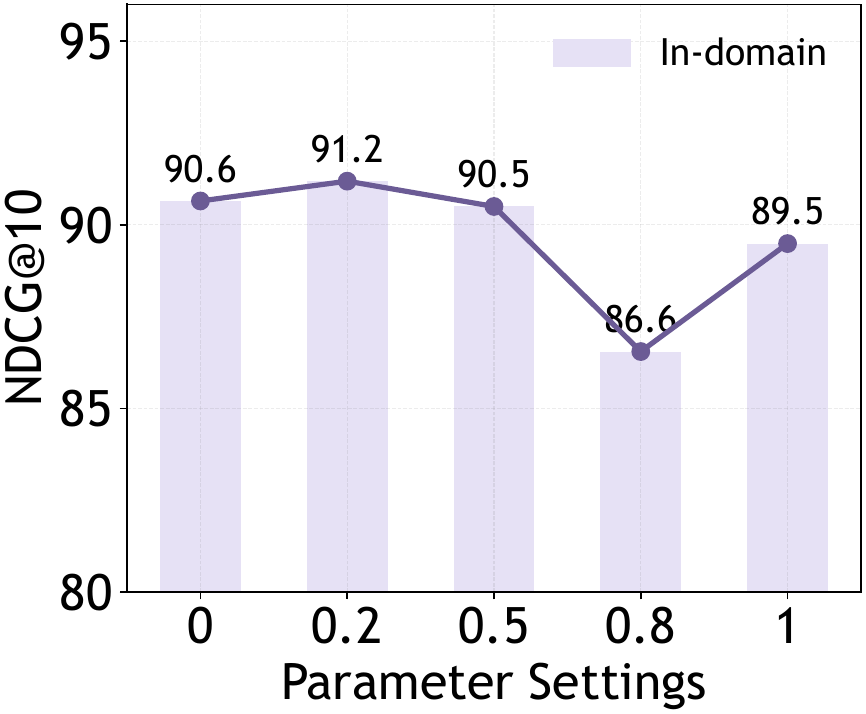}
    \caption{In-Domain}
\end{subfigure}
\begin{subfigure}{0.24\textwidth}
    \centering
    \includegraphics[width=\linewidth]{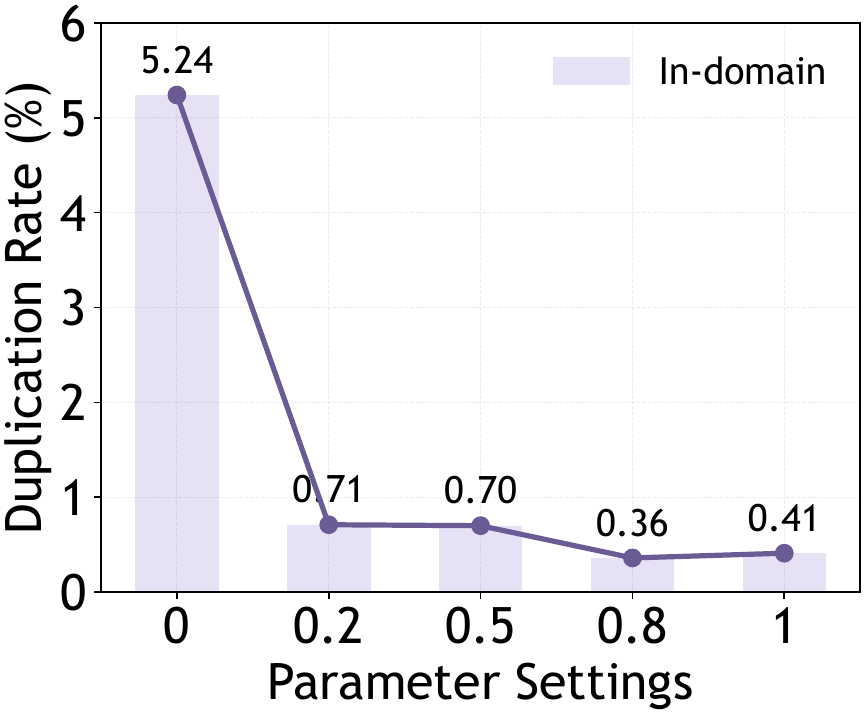}
     \caption{In-Domain}
\end{subfigure}
\begin{subfigure}{0.24\textwidth}
    \centering
    \includegraphics[width=\linewidth]{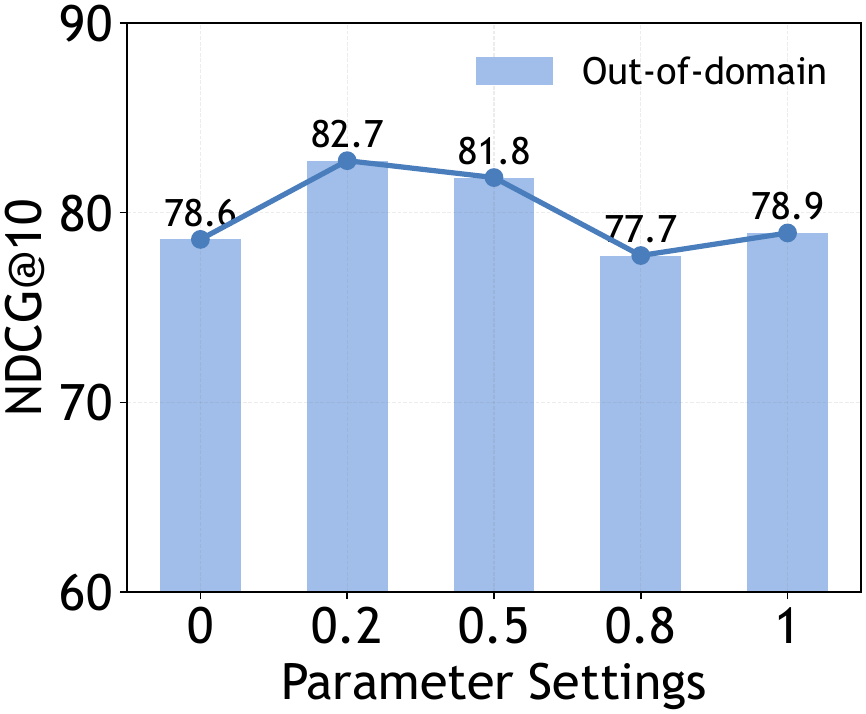}
     \caption{Out-of-Domain}
\end{subfigure}
\begin{subfigure}{0.24\textwidth}
    \centering
    \includegraphics[width=\linewidth]{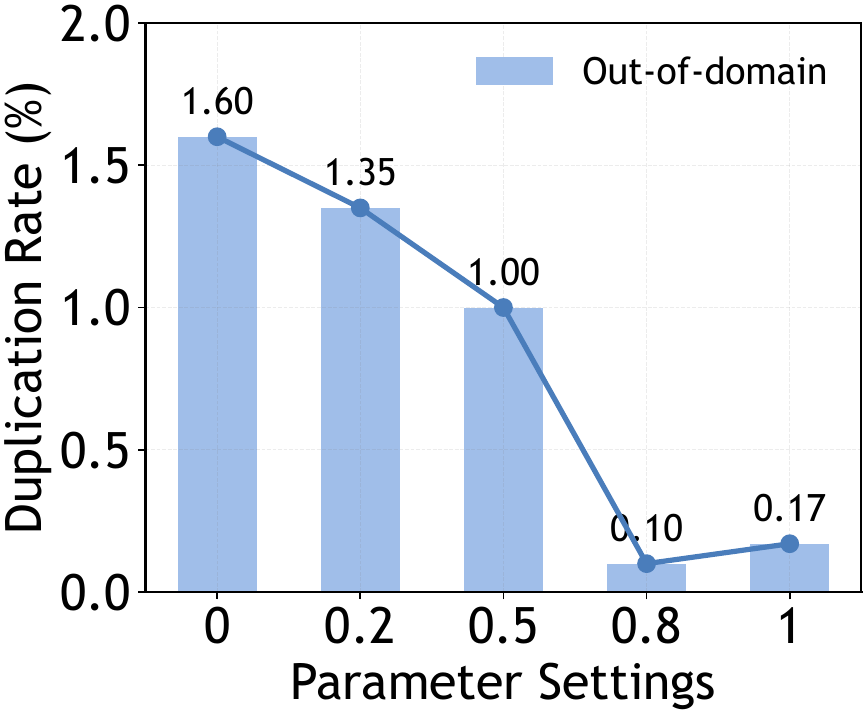}
     \caption{Out-of-Domain}
\end{subfigure}
\caption{Hyperparameter analysis under different $\alpha$ settings.}
\label{fig:hyperparameter_analysis}
\vspace{-13pt}
\end{figure*}

\textbf{When might MagicSelector fail?} We further analyze that improper reward weight settings can lead to the failure of MagicSelector. As shown in Figure~\ref{fig:hyperparameter_analysis} (a)-(c), when the preference reward weight $\alpha=0$, the model relies solely on counterfactual signals for optimization. Although it maintains high retrieval accuracy in-domain (N@10 = 90.6), the duplication  rate surges to 5.24\%, and out-of-domain performance drops sharply to 78.6. This confirms that pure causal optimization without structural constraints still induces reward hacking, resulting in generalization failure. Conversely, an excessively high $\alpha$ may slightly sacrifice retrieval accuracy due to over-regularization, despite suppressing redundancy. Therefore, the risk of MagicSelector failure primarily stems from extreme imbalances in reward weights; only by maintaining a dynamic balance between causal gain and structural constraints can dual optimality in retrieval accuracy and generalization robustness be achieved.

\section{Conclusion}
In this work, we present MagicSelector, a joint optimization framework specifically designed to address the fundamental challenges of tool retrieval and task planning in Large Language Model (LLM) agents. A counterfactual task decomposition mechanism, a progressive reranking method, and a dynamic Top-K strategy collectively contribute to the framework's capability to translate ambiguous user instructions into executable atomic subtasks. Specifically, we introduced a preference-guided counterfactual task decomposition mechanism that utilizes counterfactual rewards to cut off spurious correlations and impose fine-grained structural supervision on logical coherence. Furthermore, we designed a progressive tool reranking method driven by self-distillation hard negative mining to enhance fine-grained discrimination among highly similar tools. Finally, we implemented a dual semantic boundary-aware dynamic Top-K strategy that adaptively monitors reranking score cliffs and inter-tool semantic shifts to maximize relevant tool recall while filtering long-tail noise. Empirical results on MTDTool, our newly constructed benchmark tailored for mobile multi-turn interactions, demonstrate that MagicSelector significantly outperforms state-of-the-art methods in terms of tool retrieval accuracy, out-of-domain (OOD) generalization capability, and overall token efficiency. We believe this work not only provides a robust solution for high-precision tool retrieval in complex mobile assistant scenarios but also offers a scalable and effective paradigm for developing autonomous agents capable of navigating ambiguous, open-ended real-world environments.

~\\
\noindent \textbf{\large Author Contributions} \\[1em]
\textbf{Project Leader} \\[0.5em]
Zhengzong Chen, Yuanyuan Zhao, Fei Huang \\[1em]
\textbf{Core Contributors} \\[0.5em]
Lei Tang, Lijun Liu, Chuandi Jiang, Fan Yang, Keyun Chu, Chu Zhao, Shihao Liu, Minghang Li \\[1em]
\textbf{Contributors} \\[0.5em]
Bo Liang, Can Wen, Hailong Wu, Jingnan Ju, Mian Liu, Nengbin Zhang, Peiqiang Wang, Penghe Nie, Qinhui Gu, Sijia Lv, Siqi Chen, Wei Zhang, Yang Xu, Yuhao Qian, Yuxiang Zhang, Zeng Cheng, Zhen Wang, Zuan Chen

\clearpage
\bibliography{2026_conference}
\bibliographystyle{2026_conference}

\clearpage
\appendix
\startcontents[appendix]

\section*{Appendix}
\addcontentsline{toc}{section}{Appendix}

\begin{center}
    \large\bfseries Contents
\end{center}

\printcontents[appendix]{l}{1}{
    \setcounter{tocdepth}{2}
}


\begin{figure*}[t]
\section{Tool Schemas and Prompt Templates}
\label{baseline}
\vspace{1em}


\centering
\begin{minipage}{0.95\linewidth}
\hrule
\vspace{0.5em}
\noindent\textbf{System:} You are a real human phone user interacting with an AI voice assistant. Generate a natural spoken command.
\vspace{0.5em}

\noindent\textbf{User:}
\begin{quote}
\noindent Below is the list of tools available to the assistant (for your reference only; you must trigger them implicitly):
\begin{verbatim}
available_tools:
[
{"tool_name": "create_reminder",
"description": "Create, query, delete and modify reminders",
"actions": ["create", "query", "delete"],
"objects": ["time", "title", "repeat rule"]},
{"tool_name": "weather_service",
"description": "Query weather for a specified city and date",
"actions": ["query weather"],
"objects": ["city", "date"]},
{"tool_name": "calendar_service",
"description": "Query, create, modify and delete calendar events",
"actions": ["query", "create", "delete"],
"objects": ["title", "start time", "end time", "location"]}
]
\end{verbatim}

\noindent Conceive a natural human voice command, strictly following these rules:

\noindent (1) \textbf{Order}: express demands in the exact order of the tool list.

\noindent (2) \textbf{One-tool-one-request}: exactly 3 tools $\Rightarrow$ exactly 3 demands.

\noindent (3) \textbf{Colloquial conversion}: do not copy official terminology verbatim.

\noindent (4) \textbf{No long sentences}: do not cram all demands into one sentence.

\vspace{0.3em}
\noindent\textbf{State guidance} (injected when a tool is assigned the \textit{Ask-Detail} state):
\begin{quote}
Tool \texttt{weather\_service} is in state \textbf{Ask-Detail}. You must \textbf{deliberately omit exactly one} critical information slot (e.g., city) in the corresponding demand. The omission must be achieved through natural colloquial phrasing, with no explanatory language.
\end{quote}

\noindent\textbf{Linguistic style} (one feature randomly sampled per turn):
\begin{quote}
Use \textit{fuzzy quantifiers and relative time}---avoid exact numbers or absolute timestamps; use expressions like "a bit", "later", "in a couple of days" in exactly one demand.
\end{quote}
\end{quote}

\vspace{0.5em}
\noindent\textbf{Expected output:} Set a reminder for me to pick up the kid this afternoon. Oh, by the way, what's the weather like tomorrow? And while you're at it, check if I've got any meetings on Wednesday.
\vspace{0.3em}
\hrule
\end{minipage}
\caption*{First-turn user prompt for instruction generation. Variable blocks (tool list, state guidance, linguistic style) are dynamically injected based on the state-machine configuration.}
\label{ap:user_prompt}
\end{figure*}







\begin{figure*}[htbp]
\centering

\begin{minipage}{0.95\linewidth}
\small 
\hrule
\vspace{0.2em} 
\noindent\textbf{System:} You are an AI voice assistant. Generate a natural, helpful response based on the user's command and backend execution states.
\vspace{0.2em} 

\noindent\textbf{User:} ``Set a reminder for me to pick up the kid this afternoon. Oh, by the way, what's the weather like tomorrow? And while you're at it, check if I've got any meetings on Wednesday.''
\vspace{0.2em} 

\noindent\textbf{Assistant:}
\vspace{-0.5em} 
\begin{quote}
\noindent\textit{Backend execution states:}
\vspace{-0.3em} 
\begin{lstlisting}[breaklines=true,basicstyle=\ttfamily\small]
- Tool: Create, query, delete and modify reminders
 State: [Success]
- Tool: Query weather for a specified city and date
 State: [Ask-Detail]
 Follow-up direction: city
- Tool: Query, create, modify and delete calendar events
 State: [Silent]
\end{lstlisting}
\vspace{-0.3em} 

\noindent (2) \textbf{Coherent delivery}: blend multiple task feedback into one fluent response; no bullet-point reporting.

\noindent (3) \textbf{No tool leakage}: never reveal tool names, codes, or backend terminology.
\end{quote}
\vspace{-0.2em} 

\vspace{0.1em} 
\noindent\textbf{Expected output:} ``Your reminder to pick up the kid this afternoon has been set. By the way, which city would you like me to check the weather for tomorrow?''
\vspace{0.2em} 
\hrule
\end{minipage}
\caption*{First-turn assistant prompt. Responses are strictly driven by execution states; the \textit{Silent} tool is completely absent from the output.}
\label{ap:assistant_prompt_example}

\vspace{0.5em} 

\begin{minipage}{0.95\linewidth}
\small 
\hrule
\vspace{0.2em} 
\noindent\textbf{System:} You are a real human phone user. Generate the next-turn spoken command, building on the prior conversation.
\vspace{0.2em} 

\noindent\textit{Dialogue history:}
\vspace{-0.3em} 
\begin{quote}
\textbf{User:} Set a reminder to pick up the kid this afternoon. Oh, by the way, what's the weather like tomorrow? And check if I've got any meetings on Wednesday.''\\ \textbf{Assistant:} Your afternoon reminder is set. Which city would you like me to check the weather for?''
\end{quote}
\vspace{-0.3em} 

\noindent\textbf{User (next turn):}
\vspace{-0.5em} 
\begin{quote}
\noindent\textit{This turn requires:}
\vspace{-0.3em} 
\begin{itemize}[nosep,leftmargin=*]
\item Answer the assistant's follow-up question about the missing city.
\item Introduce new demands corresponding to the tools below.
\end{itemize}
\vspace{0.2em}

\noindent\textit{Tools and target states:}
\vspace{-0.3em} 
\begin{lstlisting}[breaklines=true,basicstyle=\ttfamily\small]
["weather_service", "calendar_service", "wifi_service"]
States: [Ask-Detail, Select, Silent]
\end{lstlisting}
\vspace{-0.3em} 

\noindent\textbf{Rules:}
\vspace{-0.3em} 
\begin{enumerate}[nosep,leftmargin=*]
\item \textbf{Supplement}: provide the previously omitted city name naturally; do not repeat the full original instruction.
\item \textbf{Selection}: choose one of the assistant's offered options; output only the option itself.
\item \textbf{Silent}: completely skip this tool---no mention, hint, or summary.
\item \textbf{Order}: demands must follow the tool list order.
\item \textbf{One-tool-one-request}: exactly 3 tools $\Rightarrow$ 3 demands.
\item \textbf{No confirmation words}: no "okay", "thanks", etc.
\end{enumerate}
\vspace{0.2em}

\noindent\textit{Linguistic style (one demand only):} Use \textit{inverted sentence}---front-load the object before the action, e.g., ``That QQ Music, open it for me.''
\end{quote}
\vspace{-0.2em} 

\vspace{0.1em} 
\noindent\textbf{Expected output:} ``Beijing. The first meeting on Wednesday, cancel that. And turn on the Wi-Fi.''
\vspace{0.2em} 
\hrule
\end{minipage}
\caption*{Next-turn user prompt (context-inheritance mode). The user must simultaneously answer the assistant's prior follow-up, make a selection, introduce new demands, and silently skip one tool---all in a single utterance.}

\end{figure*}

\begin{figure*}[htbp]
\centering

\begin{minipage}{0.95\linewidth}
\small 
\hrule
\vspace{0.2em} 
\noindent\textbf{System:} You are an AI voice assistant. Generate a natural response based on the user's next-turn command and backend execution states.
\vspace{0.2em} 

\noindent\textbf{User:} ``Beijing. The first meeting on Wednesday, cancel that. And turn on the Wi-Fi.''
\vspace{0.2em} 

\noindent\textbf{Assistant:}
\vspace{-0.5em} 
\begin{quote}
\noindent\textit{Backend execution states:}
\vspace{-0.3em} 
\begin{lstlisting}[breaklines=true,basicstyle=\ttfamily\small]
Tool: Query weather for a specified city and date State: [Success]
Tool: Query, create, modify and delete calendar events State: [Ask-Detail] Follow-up direction: specific date to cancel
Tool: Toggle Wi-Fi connectivity State: [Executing] 
\end{lstlisting}
\vspace{-0.3em} 

\noindent\textbf{Response rules:}
\vspace{-0.3em} 
\begin{enumerate}[nosep,leftmargin=*]
\item \textbf{State-driven}:
\begin{itemize}[nosep]
\item \textit{Success}: confirm completion.
\item \textit{Ask-Detail}: ask a specific question for the missing information.
\item \textit{Executing}: inform the user the task is being processed.
\item \textit{Silent}: completely skip, as if the demand never existed.
\end{itemize}
\item \textbf{Coherent}: blend all feedback into one fluent response.
\item \textbf{No tool leakage}: never reveal tool names or backend terms.
\end{enumerate}
\end{quote}
\vspace{-0.2em} 

\vspace{0.1em} 
\noindent\textbf{Expected output:} ``I've checked the weather for Beijing---it looks clear tomorrow. Which specific meeting on Wednesday would you like me to cancel? Meanwhile, I'm turning on the Wi-Fi for you.''
\vspace{0.2em} 
\hrule
\end{minipage}
\caption*{Next-turn assistant prompt. Identical rule structure to the first-turn version; only the injected user query and tool states differ.}

\vspace{0.5em} 

\begin{minipage}{0.95\linewidth}
\small 
\hrule
\vspace{0.2em} 
\noindent\textbf{System:} You are a multi-turn rewriting and task decomposition expert. Decompose user commands into single-turn, single-task units.
\vspace{0.2em} 

\noindent\textit{Dialogue history:}
\vspace{-0.3em} 
\begin{quote}
\textbf{User:} Set a reminder to pick up the kid this afternoon. Oh, by the way, what's the weather like tomorrow? And check if I've got any meetings on Wednesday.''\\ \textbf{Assistant:} Your afternoon reminder is set. Which city would you like me to check the weather for?''
\end{quote}
\vspace{-0.3em} 

\noindent\textbf{Current query:} ``Beijing. The first meeting on Wednesday, cancel that. And turn on the Wi-Fi.''
\vspace{0.2em} 

\noindent\textbf{Task definition:}
\vspace{-0.3em} 
\begin{enumerate}[nosep,leftmargin=*]
\item \textbf{Rewriting}: resolve pronouns and omissions using context. E.g., (Q: open Bluetooth)(A: done)(Q: turn it off)'' $\rightarrow$ turn off Bluetooth''.
\item \textbf{Multi-task splitting}: separate compound commands into atomic tasks. E.g., (Q: open Wi-Fi, Bluetooth and dark mode)'' $\rightarrow$ open Wi-Fi$<br>$open Bluetooth$<br>$enable dark mode''.
\end{enumerate}
\vspace{0.2em} 

\noindent\textbf{Output format:} separate tasks with \texttt{<br>}.

\vspace{0.2em} 
\noindent\textbf{Expected output:}
\vspace{-0.5em} 
\begin{quote}
``Query weather for Beijing$<br>$cancel the first Wednesday meeting$<br>$turn on Wi-Fi''
\end{quote}
\vspace{-0.2em} 

\vspace{0.1em} 
\hrule
\end{minipage}
\caption*{Task decomposition prompt. Invoked independently at each turn to produce an atomic task sequence $\mathcal{A}_t$, forming the middle tier of the three-level annotation structure (context $\rightarrow$ atomic tasks $\rightarrow$ candidate tools).}
\end{figure*}

\clearpage
\section{Case Studies}
\label{baseline}

\vspace{1em}
\begin{center}

\begin{minipage}{0.98\textwidth}
\hrule
\vspace{0.2em}
\noindent\textbf{\large Part 1: Multi-turn Dialogue}
\vspace{0.2em}
\hrule
\vspace{0.2em}

\noindent\small\textbf{Round 1 (Initial)} --- Selected Tools: 
\texttt{check\_weather}, \texttt{wifi\_service} \\
States: [Ask-Detail, Executing]

\vspace{0.1em}
\noindent\textbf{User:} 
``What's the weather like tomorrow? And turn on the Wi-Fi.''

\vspace{0.1em}
\noindent\textbf{Task split:} \\
``Query tomorrow's weather'' \\
``Turn on Wi-Fi''

\vspace{0.1em}
\noindent\textbf{Retrieval \& Reranking:} \\
\begin{tabular}{@{}l p{0.75\textwidth}@{}}
Task 1 Candidates: & [\texttt{check\_weather}, \texttt{search\_location}, \texttt{hotel\_service}, \texttt{browser\_service}, \texttt{location\_service}] \\
Task 2 Candidates: & [\texttt{wifi\_service}, \texttt{network\_settings}, \texttt{device\_control}, \texttt{system\_settings}, \texttt{bluetooth\_service}]
\end{tabular}

\vspace{0.1em}
\noindent\textbf{Tool Orchestration:} \\
\texttt{check\_weather}(time="tomorrow") $\parallel$ \texttt{wifi\_service}(action="turn\_on")

\vspace{0.1em}
\noindent\textbf{Assistant:} 
``Which city would you like me to check the weather for tomorrow? 
Meanwhile, I'm enabling Wi-Fi for you.''

\vspace{0.3em}

\noindent\small\textbf{Round 2 (Inherit)} --- Selected Tools: 
\texttt{check\_weather}, \texttt{create\_reminder} \\
States: [Success, Success]

\vspace{0.1em}
\noindent\textbf{User:} 
``Beijing. Also remind me to bring my laptop tomorrow.''

\vspace{0.1em}
\noindent\textbf{Task split:} \\
``Query weather for Beijing'' \\
``Remind to bring laptop tomorrow''

\vspace{0.1em}
\noindent\textbf{Retrieval \& Reranking:} \\
\begin{tabular}{@{}l p{0.75\textwidth}@{}}
Task 1 Candidates: & (Inherited from Round 1) \\
Task 2 Candidates: & [\texttt{create\_reminder}, \texttt{open\_reminder}, \texttt{search\_service}, \texttt{note\_service}, \texttt{system\_settings}]
\end{tabular}

\vspace{0.1em}
\noindent\textbf{Tool Orchestration:} \\
\texttt{check\_weather}(location="Beijing", time="tomorrow") $\parallel$ \texttt{create\_reminder}(content="bring laptop", time="tomorrow")

\vspace{0.1em}
\noindent\textbf{Assistant:} 
``Got it---tomorrow in Beijing will be partly cloudy, around 18 degrees. 
I have also set a reminder for you to bring your laptop tomorrow.''

\vspace{0.3em}

\noindent\small\textbf{Round 3 (Switch)} --- Selected Tools: 
\texttt{music\_service}, \texttt{bluetooth\_service} \\
States: [Success, Success]

\vspace{0.1em}
\noindent\textbf{User:} 
``Play some light music. Oh, and connect the headphones.''

\vspace{0.1em}
\noindent\textbf{Task split:} \\
``Play light music'' \\
``Connect Bluetooth headphones''

\vspace{0.1em}
\noindent\textbf{Retrieval \& Reranking:} \\
\begin{tabular}{@{}l p{0.75\textwidth}@{}}
Task 1 Candidates: & [\texttt{music\_service}, \texttt{search\_media}, \texttt{query\_media}, \texttt{audio\_control}, \texttt{video\_service}] \\
Task 2 Candidates: & [\texttt{bluetooth\_service}, \texttt{device\_control}, \texttt{wifi\_service}, \texttt{network\_settings}, \texttt{system\_settings}]
\end{tabular}

\vspace{0.1em}
\noindent\textbf{Tool Orchestration:} \\
\texttt{music\_service}(genre="light music", action="play") $\parallel$ \texttt{bluetooth\_service}(device="headphones", action="connect")

\vspace{0.1em}
\noindent\textbf{Assistant:} 
``Playing a light music playlist. 
Your Bluetooth headphones are now connected.''

\vspace{0.2em}
\hrule
\vspace{0.2em}
\captionof*{figure}{Case 1: Multi-turn Dialogue. Round~1 initializes states; Round~2 inherits context to answer follow-ups while introducing a new task; Round~3 switches to an entirely new topic.}
\label{ap:multi_turn_prompt}
\end{minipage}

\vspace{1em}

\begin{minipage}{0.98\textwidth}
\hrule
\vspace{0.4em}
\noindent\textbf{\large Part 2: Single-turn Multi-task}
\vspace{0.4em}
\hrule
\vspace{0.4em}

\noindent\small Selected Tools: 
\texttt{create\_reminder}, \texttt{check\_weather} \\
States: [Success, Success]

\vspace{0.3em}
\noindent\textbf{User:} ``Set a reminder to pick up the kid this afternoon. 
Oh, by the way, what's the weather like in Beijing tomorrow?''

\vspace{0.2em}
\noindent\textbf{Task split:} \\
``Set reminder to pick up kid this afternoon'' \\
``Query tomorrow's weather in Beijing''

\vspace{0.2em}
\noindent\textbf{Retrieval \& Reranking:} \\
\begin{tabular}{@{}l p{0.75\textwidth}@{}}
Task 1 Candidates: & [\texttt{create\_reminder}, \texttt{search\_service}, \texttt{open\_reminder}, \texttt{search\_media}, \texttt{query\_media}] \\
Task 2 Candidates: & [\texttt{check\_weather}, \texttt{search\_location}, \texttt{hotel\_service}, \texttt{browser\_service}, \texttt{location\_service}]
\end{tabular}

\vspace{0.2em}
\noindent\textbf{Tool Orchestration:} \\
\texttt{create\_reminder}(content="pick up kid", time="this afternoon") $\parallel$ \texttt{check\_weather}(location="Beijing", time="tomorrow")

\vspace{0.2em}
\noindent\textbf{Assistant:} 
``Your afternoon reminder is set. 
Tomorrow in Beijing will be partly cloudy, around 18 degrees.''

\vspace{0.4em}
\hrule
\vspace{0.4em}
\captionof*{figure}{Case 2: Single-turn Multi-task. The user issues multiple independent intents in a single utterance. The system decomposes the query, performs retrieval and reranking for each sub-task, orchestrates the selected tools, and generates a combined response.}
\end{minipage}

\end{center}
\vspace{1em}

\end{document}